\documentclass[sigconf]{acmart}

\usepackage{bbding}
\usepackage{multirow}
\usepackage{diagbox}
\usepackage{bm}
\usepackage{subfigure}
\usepackage{url}

\AtBeginDocument{%
  \providecommand\BibTeX{{%
    \normalfont B\kern-0.5em{\scshape i\kern-0.25em b}\kern-0.8em\TeX}}}

\copyrightyear{2022} 
\acmYear{2022} 
\setcopyright{acmcopyright}\acmConference[MM '22]{Proceedings of the 30th ACM International Conference on Multimedia}{October 10--14, 2022}{Lisboa, Portugal}
\acmBooktitle{Proceedings of the 30th ACM International Conference on Multimedia (MM '22), October 10--14, 2022, Lisboa, Portugal}
\acmPrice{15.00}
\acmDOI{10.1145/3503161.3548200}
\acmISBN{978-1-4503-9203-7/22/10}

\begin{document}

\title{Where Are You Looking?: A Large-Scale Dataset of Head and Gaze Behavior for 360-Degree Videos and a Pilot Study}

%%
%%% The "author" command and its associated commands are used to define
%%% the authors and their affiliations.
%%% Of note is the shared affiliation of the first two authors, and the
%%% "authornote" and "authornotemark" commands
%%% used to denote shared contribution to the research.
%\email{webmaster@marysville-ohio.com}
%\affiliation{%
%  \institution{Institute for Clarity in Documentation}
%  \streetaddress{P.O. Box 1212}
%  \country{USA}
\author{Yili Jin}
\authornotemark[2]
\affiliation{%
  \institution{The Chinese University of Hong Kong, Shenzhen}
  \city{Shenzhen}
  \country{China}}
\email{yilijin@link.cuhk.edu.cn}

\author{Junhua Liu}
\authornotemark[2]
\affiliation{%
  \institution{The Chinese University of Hong Kong, Shenzhen}
  \city{Shenzhen}
  \country{China}}
\email{junhualiu@link.cuhk.edu.cn}

\author{Fangxin Wang}
\authornote{Fangxin Wang is the corresponding author.\\$\dagger$Both authors contributed equally to this research.}
\affiliation{%
  \institution{SSE and FNii, The Chinese University of Hong Kong, Shenzhen\\Peng Cheng Laboratory}
  \city{Shenzhen}
  \country{China}}
\email{wangfangxin@cuhk.edu.cn}

\author{Shuguang Cui}
\affiliation{%
  \institution{SSE and FNii, The Chinese University of Hong Kong, Shenzhen\\Shenzhen Research Institute of Big Data\\Peng Cheng Laboratory}
  \city{Shenzhen}
  \country{China}}
\email{shuguangcui@cuhk.edu.cn}
%\affiliation{%
%  \institution{Inria Paris-Rocquencourt}
%\affiliation{%
% \institution{Rajiv Gandhi University}
%\affiliation{%
%  \institution{Tsinghua University}
%  \streetaddress{30 Shuangqing Rd}
%\affiliation{%
%  \institution{Palmer Research Laboratories}
%  \streetaddress{8600 Datapoint Drive}
%  \country{USA}
%\affiliation{%
%  \institution{The Th{\o}rv{\"a}ld Group}
%  \streetaddress{1 Th{\o}rv{\"a}ld Circle}
%\email{jsmith@affiliation.org}
%\affiliation{%
%  \institution{The Kumquat Consortium}
%  \city{New York}
%\email{jpkumquat@consortium.net}
%
%%%
%%% By default, the full list of authors will be used in the page
%%% headers. Often, this list is too long, and will overlap
%%% other information printed in the page headers. This command allows
%%% the author to define a more concise list
%%% of authors' names for this purpose.

\begin{abstract}
360° videos in recent years have experienced booming development. Compared to traditional videos, 360° videos are featured with uncertain user behaviors, bringing opportunities as well as challenges. Datasets are necessary for researchers and developers to explore new ideas and conduct reproducible analyses for fair comparisons among different solutions. However, existing related datasets mostly focused on users' field of view (FoV), ignoring the more important eye gaze information, not to mention the integrated extraction and analysis of both FoV and eye gaze. Besides, users’ behavior patterns are highly related to videos, yet most existing datasets only contained videos with subjective and qualitative classification from video genres, which lack quantitative analysis and fail to characterize the intrinsic properties of a video scene.

To this end, we first propose a quantitative taxonomy for 360° videos that contains three objective technical metrics. Based on this taxonomy, we collect a dataset containing users' head and gaze behaviors simultaneously, which outperforms existing datasets with rich dimensions, large scale, strong diversity, and high frequency. Then we conduct a pilot study on users' behaviors and get some interesting findings such as user's head direction will follow his/her gaze direction with the most possible time interval. A case of application in tile-based 360° video streaming based on our dataset is later conducted, demonstrating a great performance improvement of existing works by leveraging our provided gaze information.

Our dataset is available at \url{https://cuhksz-inml.github.io/head_gaze_dataset/}

\end{abstract}


\ccsdesc{Human-centered computing~Virtual reality}
\ccsdesc{Information systems~Multimedia databases}

\keywords{360° videos, Dataset, User Behavior Analysis}


\maketitle

\begin{table*}[t]
\begin{tabular}{|c|c|cccc|c|c|c|c|}
\hline
\multirow{2}{*}{Dataset}                                                   & \multirow{2}{*}{Size} & \multicolumn{4}{c|}{Details of Size Information}                                                                  & \multirow{2}{*}{Content} & \multirow{2}{*}{Head}     & \multirow{2}{*}{Gaze}     & \multirow{2}{*}{\begin{tabular}[c]{@{}c@{}}Diversity of \\Contents\end{tabular}} \\ \cline{3-6}
                                                                           &                       & \multicolumn{1}{c|}{Length (s)} & \multicolumn{1}{c|}{No. Contents} & \multicolumn{1}{c|}{No. Users} & Freq. (Hz) &                          &                           &                           &                                                                                    \\ \hline
Corbillon et al. \cite{DBLP:conf/mmsys/CorbillonSS17}     & Tiny                 & \multicolumn{1}{c|}{70}         & \multicolumn{1}{c|}{5}            & \multicolumn{1}{c|}{59}        & 30         & Video                    & \checkmark & --                        & --                      \\ \hline
Fremerey et al. \cite{DBLP:conf/mmsys/FremereySMR18}      & Tiny                 & \multicolumn{1}{c|}{30}         & \multicolumn{1}{c|}{20}           & \multicolumn{1}{c|}{48}        & 10         & Video                    & \checkmark & --                        & --                      \\ \hline
Lo et al. \cite{DBLP:conf/mmsys/LoFLHCH17}                & Small                 & \multicolumn{1}{c|}{60}         & \multicolumn{1}{c|}{10}           & \multicolumn{1}{c|}{50}        & 30         & Video                    & \checkmark & --                        & --                      \\ \hline
Nasrabadi et al. \cite{DBLP:conf/mmsys/NasrabadiSMMPFC19} & Middle                & \multicolumn{1}{c|}{60}         & \multicolumn{1}{c|}{28}           & \multicolumn{1}{c|}{60}        & 50         & Video                    & \checkmark & --                        & \checkmark                      \\ \hline
Wu et al. \cite{DBLP:conf/mmsys/WuTWY17}                  & Large                 & \multicolumn{1}{c|}{164-655}    & \multicolumn{1}{c|}{18}           & \multicolumn{1}{c|}{48}        & 100        & Video                    & \checkmark & --                        & --                      \\ \hline
Agtzidis et al. \cite{DBLP:conf/mm/AgtzidisSD19}          & Small                 & \multicolumn{1}{c|}{60}         & \multicolumn{1}{c|}{15}           & \multicolumn{1}{c|}{13}        & 120        & Video                    & --                        & \checkmark & --                      \\ \hline
Rai et al. \cite{DBLP:conf/mmsys/RaiGC17}                 & Middle                & \multicolumn{1}{c|}{25}         & \multicolumn{1}{c|}{60}           & \multicolumn{1}{c|}{40}        & 60         & Image                    & \checkmark & \checkmark & --                      \\ \hline \hline
\textbf{Ours}                                             & Large                 & \multicolumn{1}{c|}{60}         & \multicolumn{1}{c|}{27}           & \multicolumn{1}{c|}{100}       & 120        & Video                    & \checkmark & \checkmark & \checkmark                      \\ \hline
\end{tabular}
\caption{Existing datasets}
\vspace{-20pt}
\label{tbl:datasets}
\end{table*}

\section{Introduction}
Virtual Reality (VR) develops vigorously in recent years, empowering a new form of video watching with a fully immersive and interactive experience. As one of the most important manifestations of VR, 360° videos have attracted great attention and are widely explored in many multimedia applications, from games to education. It is also a supporting technology for the new paradigm \textit{Metaverse}. Indicated by Cisco Mobile Visual Networking Index (VNI)~\cite{cisco2018cisco}, 360° videos mobile data traffic grows nearly 12-fold from 2017 to 2022. According to a recent market research report published on Globe Newswire~\cite{global2021market}, the global market size was about \$6 billion in 2021. And the report predicts that the market will be projected to grow to \$80 billion in 2028 at a compound annual growth rate (CAGR) of 45\% in the 2021-2028 period.

Different from the fixed Field of View (FoV) in traditional videos, 360° videos have 3 degrees of freedom, which allow users to freely rotate their heads to watch the most attractive part. Many existing works on 360° videos are based on or focus on this feature. In psychology, FoVs are used to analyze users' mental activities \cite{elmezeny2018immersive, breves2020into, rupp2016effects}. In computer networking, tile-based HTTP streaming \cite{DBLP:conf/mmsys/GrafTM17, DBLP:conf/infocom/ZhangZBLSL19, DBLP:conf/www/ChopraCMC21} is highly dependent on users' FoVs, which assigns high quality for content in the FoV while low quality (or even blank content) for the rest of the video, so as to reduce the bandwidth consumption. 
% Also, video producers can optimize their videos based on users' behaviors.

Besides the FoV feature, users' eye gaze information is another important metric in 360° videos.
From the first device that could track eye movement built in 1908 \cite{huey1908psychology}, eye tracking technologies have experienced rapid development. Instead of initial invasive methods, e.g., direct mechanical contact with the cornea \cite{jacob2003eye}, current works can track the eye gaze precisely in a non-invasive way. On one hand, hardware such as RGB-D cameras \cite{DBLP:conf/huc/XiongCLZ14} and infrared cameras \cite{DBLP:conf/etra/WangBLAFOSS16} become much more powerful, lightweight, and cheap nowadays, enabling more precise raw gaze data collection. On the other hand, the rapidly developing computer vision and machine learning algorithms further empower eye gaze-related information prediction and processing~\cite{DBLP:journals/infsof/SharafiSG15, DBLP:books/lib/Holmqvist11}.

In practice, eye gaze data is more fine-grained than FoV data to indicate where the users are looking and can be well leveraged in many potential scenarios. Taking 360° video streaming as an example, video content in the FoV but far from the gaze center can be encoded in a lower bitrate, which can highly improve the transmission efficiency without losing QoE. Although eye gaze information reveals great potential, there are few datasets focusing on eye gaze recording in 360° video watching, not to mention the combined extraction for FoV and eye gaze simultaneously for further analysis. Besides, users’ behavior patterns are highly related to videos so it is important to understand the correlations therein. Yet most existing datasets ignored this and only contained videos with simple qualitative classification from video genres, without quantitative representation and analysis. Their classification methods are generally ambiguous and subjective, which do not characterize the intrinsic properties of a video scene. 

In this paper, we mainly propose a large-scale dataset containing both the head motion, which denotes FoV, and eye gaze information, and further conduct a comprehensive data analytics as well as a case application on gaze-assisted 360° video streaming. We first develop a quantitative taxonomy for 360° videos, which contains three objective technical metrics, i.e., camera motion, video quality, and the dispersion of region of interest (ROI). Based on the taxonomy, we implement an auto-taxonomizer, which can classify videos objectively. We then collect a 360° video dataset, which outperforms existing related datasets with rich dimensions (including both head and gaze), large scale (including 100 users and 27 videos), strong diversity (wide range across the taxonomy), and high frequency (120 Hz). 

Comprehensive analysis and a pilot study on users' behaviors are further conducted with some interesting findings, such as (1) user’s head direction will follow his/her gaze direction with a most possible time interval of 0.12 seconds; and (2) users explore more horizontally than vertically. 

We next implement a case of the application based on our dataset in 360° video streaming. By leveraging gaze data, the efficiency of existing streaming systems can be highly improved. From the findings we discovered, gaze data can be used to assist to predict FoV more accurately. We examine three different works which are adaptive to different situations and enhance their design with our gaze information. Experiments show that by adding gaze information, all kinds of previous methods can be improved. We also propose some other potential applications at the end.

Overall, the main contributions of our paper are as follows.
\begin{itemize}
    \item We propose a novel quantitative taxonomy for 360° videos, which contains three technical metrics, camera motion, video quality, and dispersion of region of interest (ROI).
    \item We collect a large-scale and high-diversity dataset with head and gaze behavior for 360° videos.
    \item We do a pilot study on users' behaviors and get some interesting findings.
    \item We demonstrate a use case of our dataset in 360° video streaming scenario, and propose some other potential applications as well as future works.
\end{itemize}
\vspace{-5pt}

%The rest of this paper is organized as follows. Section \ref{sec:review} reviews existing 360° video behavior datasets and Section \ref{sec:taxonomy} describes our proposed taxonomy. Section \ref{sec:dataset} shows some information about our dataset, followed by some visualizations in Section \ref{sec:vis}. Then we present a pilot study based on our dataset in Section \ref{sec:pilot} and a case of an application in 360° video streaming in Section \ref{sec:case}. Finally, Section \ref{sec:app} gives some examples of potential applications and Section \ref{sec:conclusion} concludes the paper.

\begin{table*}[t]
\scalebox{0.75}{
\begin{tabular}{|c||ccc||ccc||ccc|}
\hline
\multirow{2}{*}{\diagbox{ROI}{Camera}}      & \multicolumn{3}{c|}{Motionless}                                                            & \multicolumn{3}{c|}{Middle}                                                           & \multicolumn{3}{c|}{Moving}                                                             \\ \cline{2-10} 
                              & \multicolumn{1}{c|}{ID} & \multicolumn{1}{c|}{URL}           & Content                     & \multicolumn{1}{c|}{ID} & \multicolumn{1}{c|}{URL}          & Content                   & \multicolumn{1}{c|}{ID} & \multicolumn{1}{c|}{URL}         & Content                    \\ \hline
\multirow{3}{*}{Disperse}     & \multicolumn{1}{c|}{1}  & \multicolumn{1}{c|}{RAYdWuPnp-M}   & Stage at Core Entertainment & \multicolumn{1}{c|}{2}  & \multicolumn{1}{c|}{VpwIoyr0RMg}  & Creepy Doll Chase         & \multicolumn{1}{c|}{3}  & \multicolumn{1}{c|}{DXjPj87fgkg} & Pomerelle Mountain Resort  \\ \cline{2-10} 
                              & \multicolumn{1}{c|}{10} & \multicolumn{1}{c|}{xNN-bJQ4vI}    & Get Ready for the Drop      & \multicolumn{1}{c|}{11} & \multicolumn{1}{c|}{yVLfEHXQk08}  & Hog Rider & \multicolumn{1}{c|}{12} & \multicolumn{1}{c|}{tVsw0DvAWdE} & Heartland Motorsports Park \\ \cline{2-10} 
                              & \multicolumn{1}{c|}{19} & \multicolumn{1}{c|}{2RqjCAhkIRI}   & Still In Love (MV)          & \multicolumn{1}{c|}{20} & \multicolumn{1}{c|}{14gSxb3YoTE}  & LEGO Subway Surfers       & \multicolumn{1}{c|}{21} & \multicolumn{1}{c|}{m9EClKA1VeQ} & A London City Guided Tour  \\ \hline
\multirow{3}{*}{Middle}       & \multicolumn{1}{c|}{4}  & \multicolumn{1}{c|}{ZRFIdczxxkY}   & Tour a NYC Penthouse        & \multicolumn{1}{c|}{5}  & \multicolumn{1}{c|}{5Uf\_s5MZoRk} & Roller Coaster            & \multicolumn{1}{c|}{6}  & \multicolumn{1}{c|}{Q66f8ufanp4} & Waterslide Virtual Ride    \\ \cline{2-10} 
                              & \multicolumn{1}{c|}{13} & \multicolumn{1}{c|}{s\_hdc\_XiXiA} & Super Mario                 & \multicolumn{1}{c|}{14} & \multicolumn{1}{c|}{ZvZ7da8JBUk}  & Roller Coaster            & \multicolumn{1}{c|}{15} & \multicolumn{1}{c|}{9YJppTxIDM}  & The Ancient Greek Theatre  \\ \cline{2-10} 
                              & \multicolumn{1}{c|}{22} & \multicolumn{1}{c|}{ULixPLH-WA4}   & Closet Set Tour             & \multicolumn{1}{c|}{23} & \multicolumn{1}{c|}{jMyDqZe0z7M}  & Chariot Race              & \multicolumn{1}{c|}{24} & \multicolumn{1}{c|}{93nxeejhPkU} & Jurassic World Evolution   \\ \hline
\multirow{3}{*}{Intensive} & \multicolumn{1}{c|}{7}  & \multicolumn{1}{c|}{BEePFpC9qG8}   & Spotlight Stories           & \multicolumn{1}{c|}{8}  & \multicolumn{1}{c|}{2Lq86MKesG4}  & City Drive Tour           & \multicolumn{1}{c|}{9}  & \multicolumn{1}{c|}{6TlW1ClEBLY} & Aerial View of Beaches     \\ \cline{2-10} 
                              & \multicolumn{1}{c|}{16} & \multicolumn{1}{c|}{wczdECcwRw0}   & Experience a Virtual Raid   & \multicolumn{1}{c|}{17} & \multicolumn{1}{c|}{sqQgv3NSOjY}  & Pig Life Animation        & \multicolumn{1}{c|}{18} & \multicolumn{1}{c|}{9XR2CZi3V5k} & Drone Footage              \\ \cline{2-10} 
                              & \multicolumn{1}{c|}{25} & \multicolumn{1}{c|}{JpAdLz3iDPE}   & Cooking Battle              & \multicolumn{1}{c|}{26} & \multicolumn{1}{c|}{G-XZhKqQAHU}  & Spotlight Stories   & \multicolumn{1}{c|}{27} & \multicolumn{1}{c|}{AX4hWfyHr5g} & Wingsuit over Dubai        \\ \hline
\end{tabular}
}
\caption{Information of videos}
\vspace{-20pt}
\label{tbl:videos}
\end{table*}

\begin{table}[t]

\begin{tabular}{|c|cc|cc|}
\hline
\multirow{2}{*}{Gender}                                                               & \multicolumn{2}{c|}{Female}        & \multicolumn{2}{c|}{Male}       \\ \cline{2-5} 
                                                                                      & \multicolumn{2}{c|}{57}            & \multicolumn{2}{c|}{43}         \\ \hline\hline
\multirow{2}{*}{Age}                                                                  & \multicolumn{1}{c|}{18-21} & 22-25 & \multicolumn{1}{c|}{26-29} & $\geq 30$ \\ \cline{2-5} 
                                                                                      & \multicolumn{1}{c|}{53}    & 37    & \multicolumn{1}{c|}{7}    & 3 \\ \hline\hline
\multirow{2}{*}{\begin{tabular}[c]{@{}c@{}}Mobile VR Exp\\ (Times)\end{tabular}}      & \multicolumn{1}{c|}{Never} & 1-5   & \multicolumn{1}{c|}{6-20}  & $\geq 20$ \\ \cline{2-5} 
                                                                                      & \multicolumn{1}{c|}{30}    & 51    & \multicolumn{1}{c|}{13}     & 6  \\ \hline\hline
\multirow{2}{*}{\begin{tabular}[c]{@{}c@{}}Room Scale VR Exp.\\ (Times)\end{tabular}} & \multicolumn{1}{c|}{Never} & 1-5   & \multicolumn{1}{c|}{6-10}  & $\geq 10$ \\ \cline{2-5} 
                                                                                      & \multicolumn{1}{c|}{53}    & 34    & \multicolumn{1}{c|}{8}     & 5  \\ \hline\hline
\multirow{2}{*}{\begin{tabular}[c]{@{}c@{}}360° Video Exp.\\ (Times)\end{tabular}}     & \multicolumn{1}{c|}{Never} & 1-5   & \multicolumn{1}{c|}{6-10}  & $\geq 10$ \\ \cline{2-5} 
                                                                                      & \multicolumn{1}{c|}{41}    & 47    & \multicolumn{1}{c|}{7}     & 5  \\ \hline
\end{tabular}
\caption{Information of users}
\vspace{-20pt}
\centering
\label{tbl:users}
\end{table}

\section{Existing Datasets}
\label{sec:review}

Researchers and developers can do reproducible analysis and experiment for fair comparisons among different solutions on datasets. There exist several datasets that provide head movement tracks of users. Table \ref{tbl:datasets} shows some basic information about those datasets, where "Length" presents the length of single videos in the dataset, "No. Contents" and "No. Users" show the number of contents and users, and "Freq." presents the sampling frequency of the dataset.

Most datasets used segments about 60 seconds from the whole video, while Wu et al. \cite{DBLP:conf/mmsys/WuTWY17} used the whole video. Experiments show that users will be uncomfortable watching 360° videos for a long time, and a representative piece is enough for a video. So we follow the former idea of using 60s segments. Agtzidis et al. \cite{DBLP:conf/mm/AgtzidisSD19} gathered an eye movement dataset, but they only collected gaze information without recording head movement tracks. Rai et al. \cite{DBLP:conf/mmsys/RaiGC17} gathered a dataset with both head and gaze information, but their dataset is for 360° images.

The data sampling frequency of most datasets is about 30Hz. High frequency can provide a more precise analysis. Most datasets only have several hours of total viewing time, and Wu et al. \cite{DBLP:conf/mmsys/WuTWY17} collected about 70 hours of data with 18 videos because they used the whole videos.

Some datasets classified videos based on their genre, such as performance, film, etc. But those classification methods are ambiguous and subjective, which do not characterize the intrinsic properties of a scene. Nasrabadi et al. \cite{DBLP:conf/mmsys/NasrabadiSMMPFC19} were the first to consider the diversity of videos from a technical aspect, but their classification method was also subjective and rough.

\section{Taxonomy}
\label{sec:taxonomy}

\begin{figure}
	\centering
	\begin{minipage}[t]{0.45\linewidth}
		\centering
		\includegraphics[scale=1.4]{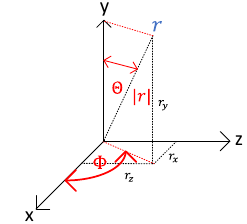}
		\caption{Sketch of $\phi$ and $\theta$}
		\vspace{-10pt}
		\label{fig:angle}
	\end{minipage}%
	\begin{minipage}[t]{0.6\linewidth}
		\centering
		\includegraphics[scale=0.4, trim = 75 30 30 65, clip]{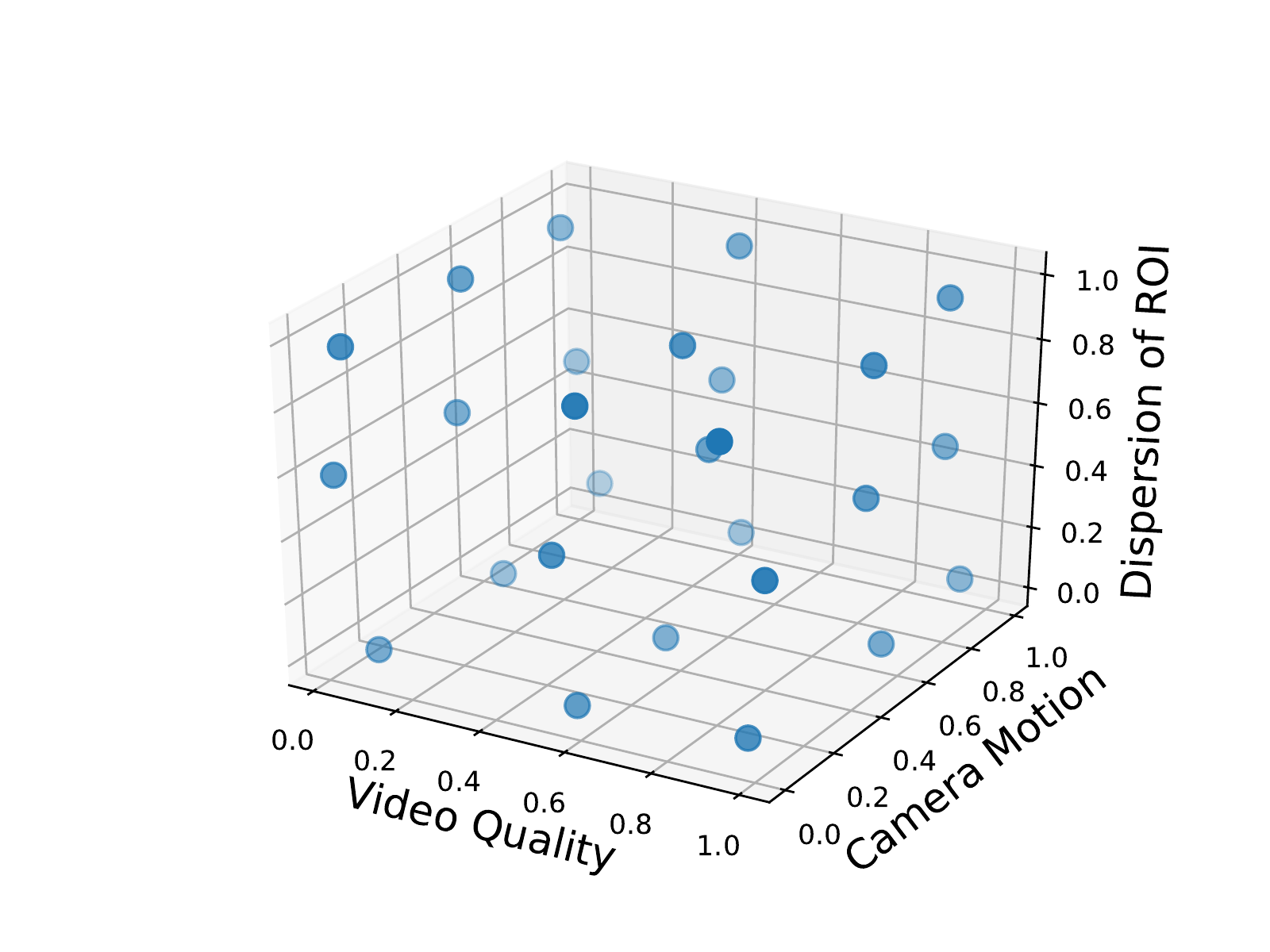}
		\caption{Diversity of videos}
		\label{fig:videos}
	\end{minipage}
\vspace{-15pt}
\end{figure}

\begin{figure*}[!t]
\normalsize
\centering
\begin{minipage}[t]{\linewidth}
\vspace{0pt}
\centering
  \hspace{-0.5cm}
  \subfigure[6'36'' video 7]{
    \label{fig:visual1}
    \includegraphics[width=0.32\linewidth, height=85pt]{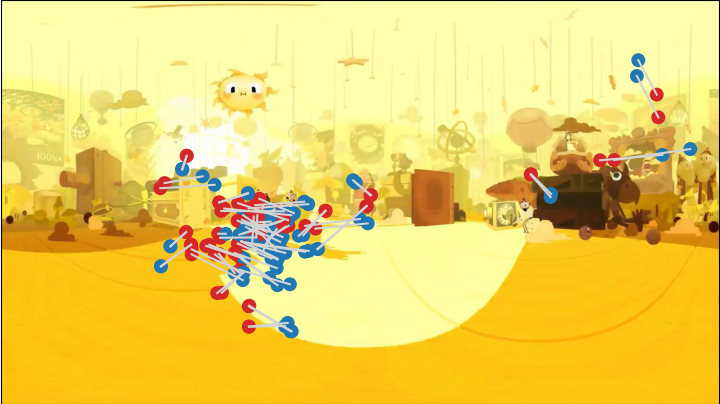}}
    \quad
    \hspace{-0.3cm}
  \subfigure[6'38'' video 7]{
    \label{fig:visual2}
    \includegraphics[width=0.32\linewidth, height=85pt]{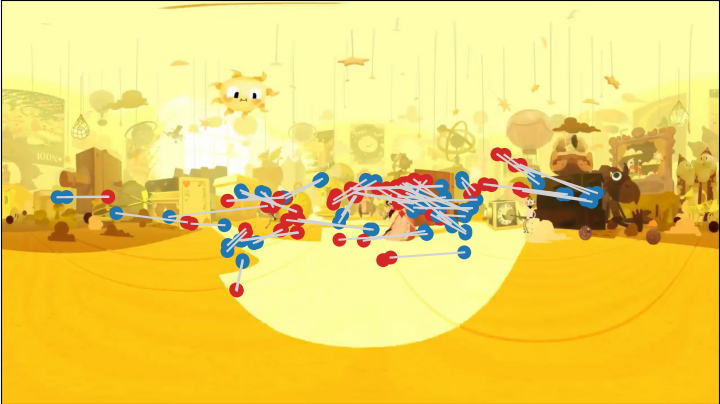}}
    \quad
    \hspace{-0.3cm}
  \subfigure[6'40'' video 7]{
    \label{fig:visual3}
    \includegraphics[width=0.32\linewidth, height=85pt]{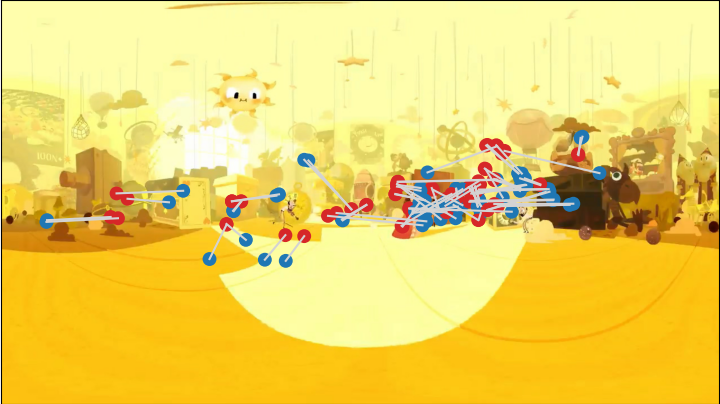}}
    \quad
    \hspace{-0.3cm}
\vspace{-0.5cm}
\caption{Users’ head and gaze direction during the motion of a moving object (Red: head direction; Blue: gaze direction)}
\label{fig:visual_1}
\end{minipage}
\end{figure*}

\begin{figure*}[!t]
\normalsize
\centering
\begin{minipage}[t]{\linewidth}
\vspace{0pt}
\centering
  \hspace{-0.5cm}
  \subfigure[1'04'' video 2]{
    \label{fig:visual4}
    \includegraphics[width=0.32\linewidth, height=85pt]{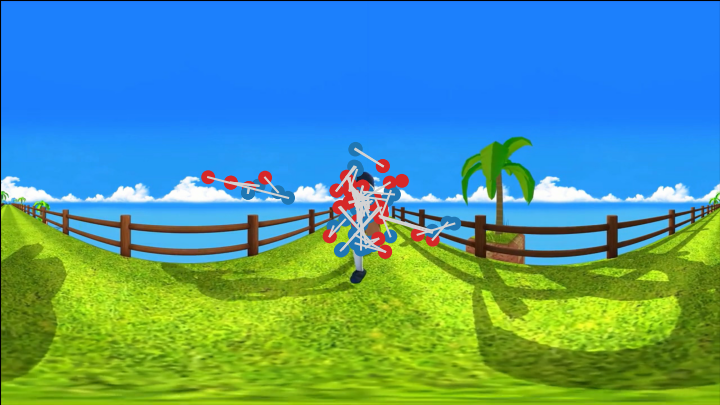}}
    \quad
    \hspace{-0.3cm}
  \subfigure[4'33'' video 5]{
    \label{fig:visual5}
    \includegraphics[width=0.32\linewidth, height=85pt]{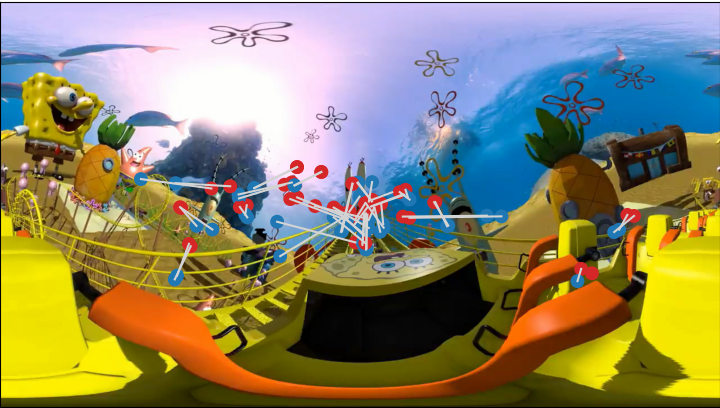}}
    \quad
    \hspace{-0.3cm}
  \subfigure[8'04'' video 9]{
    \label{fig:visual6}
    \includegraphics[width=0.32\linewidth, height=85pt]{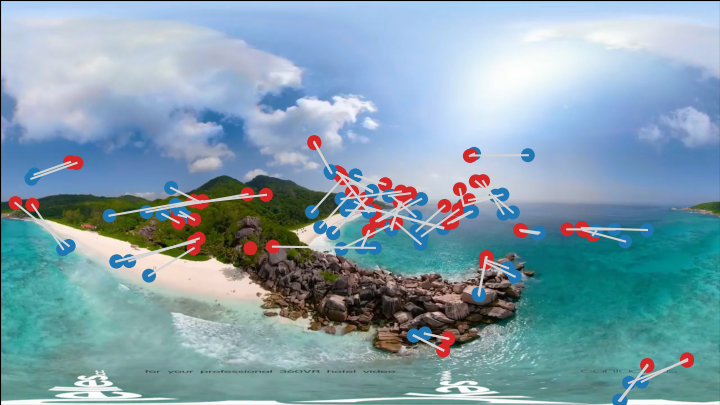}}
    \quad
    \hspace{-0.3cm}
\vspace{-0.5cm}
\caption{Users’ head and gaze direction in videos of different ROI (Red: head direction; Blue: gaze direction)}
\label{fig:visual_2}
\end{minipage}
\end{figure*}

The objective of the taxonomy is to gather videos with similar viewing patterns in the same category. We have summarized three technical metrics used as the indication, i.e., camera motion, video quality, and dispersion of region of interest (ROI). And this taxonomy is able to quantitatively represent the video features for more precise differentiation.

\subsection{Video Quality}
The quality of the video may affect users' patterns. A more clear video would encourage users to explore the environment more. The method we choose to evaluate this metric is the Weighted to Spherically uniform Peak Signal to Noise Ratio (WS-PSNR)~\cite{DBLP:journals/spl/SunLY17}.

WS-PSNR calculates PSNR using all image frames on the 2D projection plane. The distortion at each position $(i,j)$ is weighted by the spherical area covered by that sample position. For each position $(i,j)$ of an $M \times N$ image on the 2D projection plane, denote the sample values on the reference and test images as $y(i, j)$ and $y'(i , j)$ , respectively, and denote the spherical area covered by the sample as $w(i, j)$ . The weighted mean squared error (WMSE) is first calculated as:
\begin{equation}
WMSE = \frac{1}{\sum_{i=0}^{M-1}\sum_{j=0}^{N-1}w(i,j)} \sum_{i=0}^{M-1}\sum_{j=0}^{N-1}\left[ y(i, j) - y'(i , j) \right]^2 \times w(i,j)
\end{equation}

The WS-PSNR is then calculated as:
\vspace{-5pt}
\begin{equation}
WS \text{-} PSNR = 10log(\frac{MAX_I^2}{WMSE})
\end{equation}\vspace{-5pt}
where $MAX_I$ is the maximum intensity level.

\subsection{Camera Motion}
Though users can explore freely in 360° videos, the camera itself can move or rotate, which would influence users' behaviors.

There is already some research to detect the motion of the camera for traditional videos. We first cut the video into six parts from the perspective of cubic projection. Then we can leverage SfMlearner~\cite{DBLP:conf/cvpr/ZhouBSL17}, a well-known camera motion algorithm for traditional videos, to detect the camera motion from each side. Finally, we take the average of camera motions on six sides to get the final results. The motion of $n$-th side between each frame can be denoted as
\begin{equation}
    (m_{x,n},m_{y,n},m_{z,n},o_{x,n},o_{y,n},o_{z,n})
\end{equation}
where $\bf{m}$ denotes the movement and $\bf{o}$ denotes the rotation.

Then we can have the final result
\begin{equation}
    motion =  \left( \sum_{n=1}^{6} m_{x,n}, \sum_{n=1}^{6}m_{y,n}, \sum_{n=1}^{6}m_{z,n}, \sum_{n=1}^{6}o_{x,n}, \sum_{n=1}^{6}o_{y,n}, \sum_{n=1}^{6}o_{z,n} \right)
\end{equation}
\vspace{-3pt}

\subsection{Dispersion of ROI}
Users will watch the region of interest (ROI) in the video. If the ROI of the video is intensive, users' would focus on this ROI, which makes their behaviors more similar.

To calculate the dispersion of ROI, we first detect ROI by saliency map, which is an image that highlights the region on which people's eyes focus first. By finding the maximum point in the saliency map, the center of each ROI can be found. We then calculate the area of the ROI by counting from the maximum to the average. We use the area of each ROI as a weight to calculate dispersion by Standard Deviational Ellipses~\cite{10.2307/490885}.

To do this, we first get the Weighted Mean Center by
\vspace{-3pt}
\begin{equation}
\bar{X} = \frac{\sum_{i=1}^{n}w_ix_i}{\sum_{i=1}^{n}w_i}, \bar{Y} = \frac{\sum_{i=1}^{n}w_iy_i}{\sum_{i=1}^{n}w_i}
\end{equation}

Then we can get the Standard Distance, which represents dispersion, by
\vspace{-3pt}
\begin{equation}
SD = \sqrt{\frac{\sum_{i=1}^{n}w_i(x_i-\bar{X})^2}{\sum_{i=1}^{n}w_i} + \frac{\sum_{i=1}^{n}w_i(y_i-\bar{Y})^2}{\sum_{i=1}^{n}w_i}}
\end{equation}

\begin{figure*}[!t]
\normalsize
\centering
\begin{minipage}[t]{\linewidth}
\vspace{0pt}
\centering
  \hspace{-0.5cm}
  \subfigure[Video 3]{
    \label{fig:visual7}
    \centering
    \includegraphics[width=0.32\linewidth]{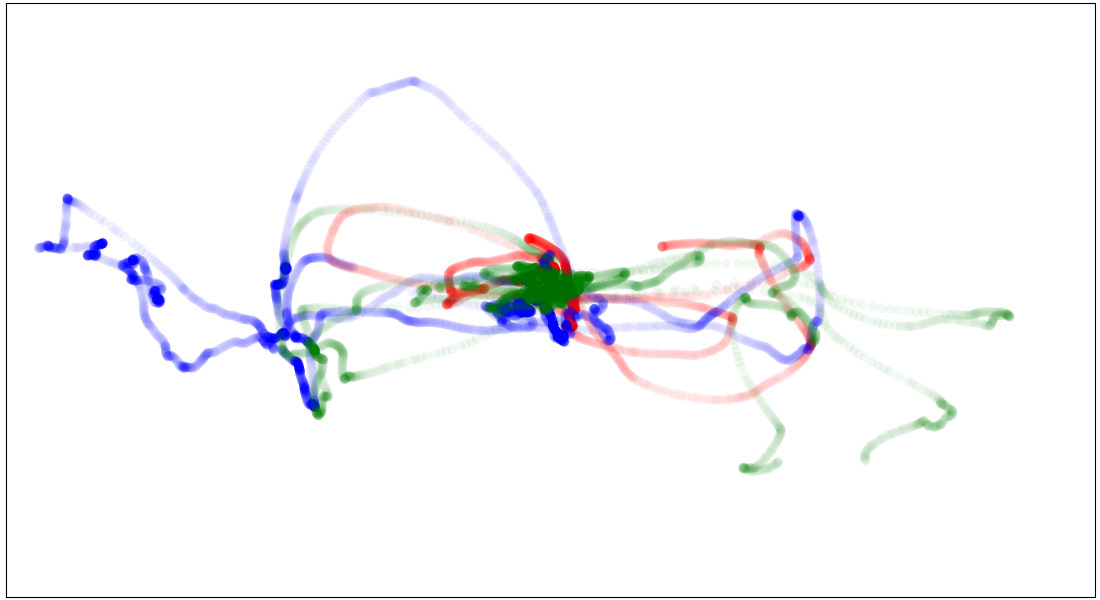}}
    \quad
    \hspace{-0.3cm}
  \subfigure[Video 6]{
    \centering
    \label{fig:visual8}
    \includegraphics[width=0.32\linewidth]{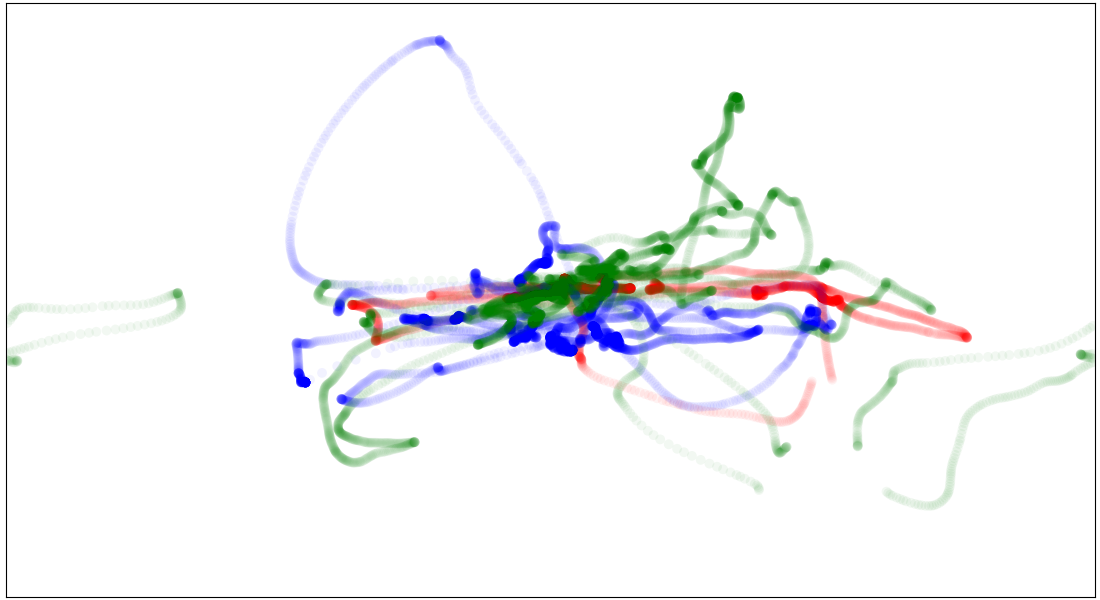}}
    \quad
    \hspace{-0.3cm}
  \subfigure[Video 9]{
    \centering
    \label{fig:visual9}
    \includegraphics[width=0.32\linewidth]{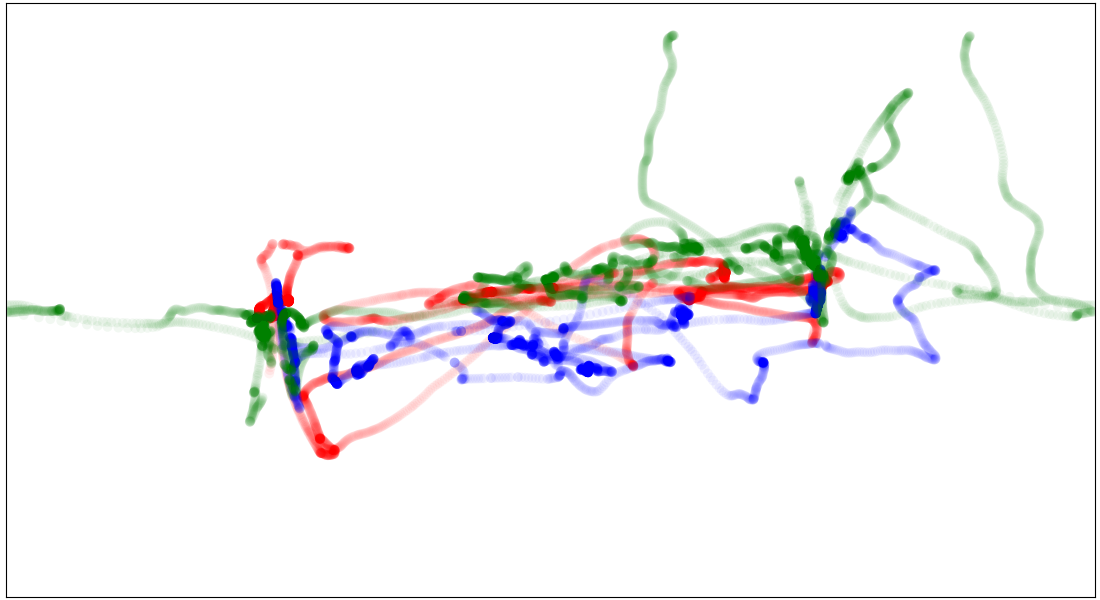}}
    \quad
    \hspace{-0.3cm}
\vspace{-0.5cm}
\caption{Density maps of three users’ head and gaze directions in different videos}
\label{fig:visual_3}
\end{minipage}
\vspace{-0.5cm}
\end{figure*}

\begin{figure*}[!t]
\normalsize
\centering
\begin{minipage}[t]{\linewidth}
\vspace{0pt}
\centering
  \hspace{-0.5cm}
  \subfigure[Video 3]{
    \label{fig:visual10}
    \centering
    \includegraphics[width=0.32\linewidth, height=115pt]{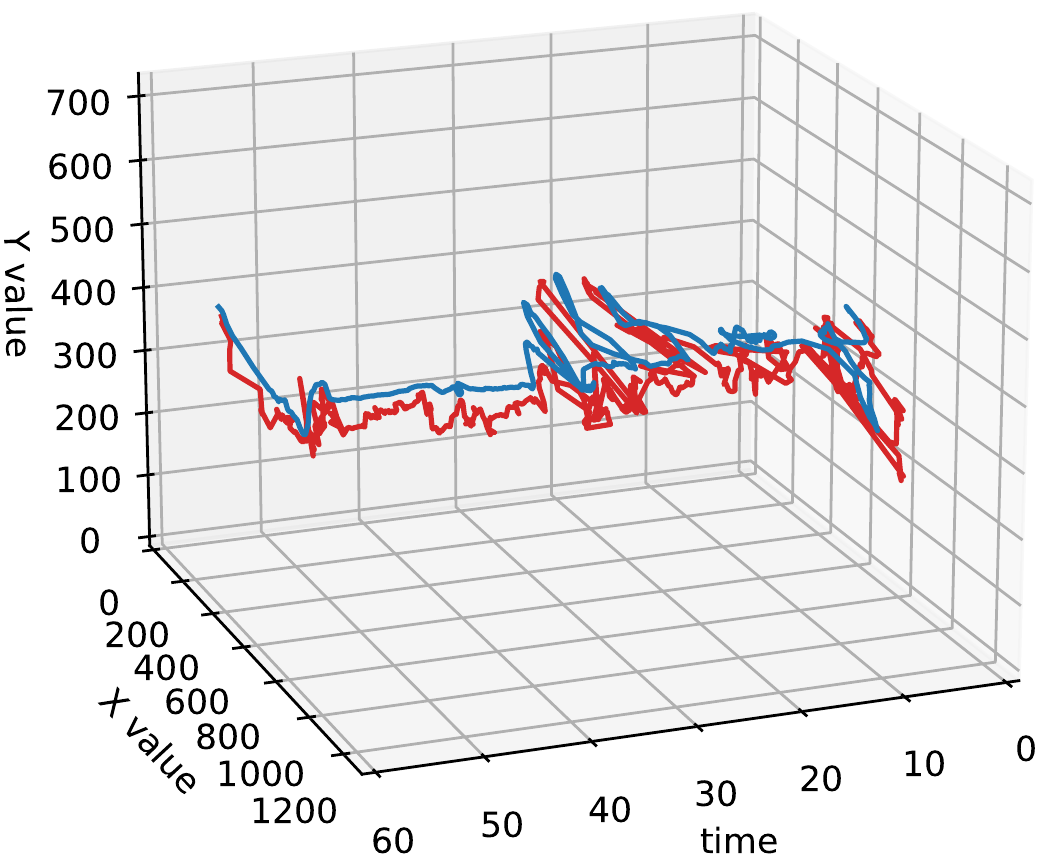}}
    \quad
    \hspace{-0.3cm}
  \subfigure[Video 6]{
    \label{fig:visual11}
    \centering
    \includegraphics[width=0.32\linewidth, height=115pt]{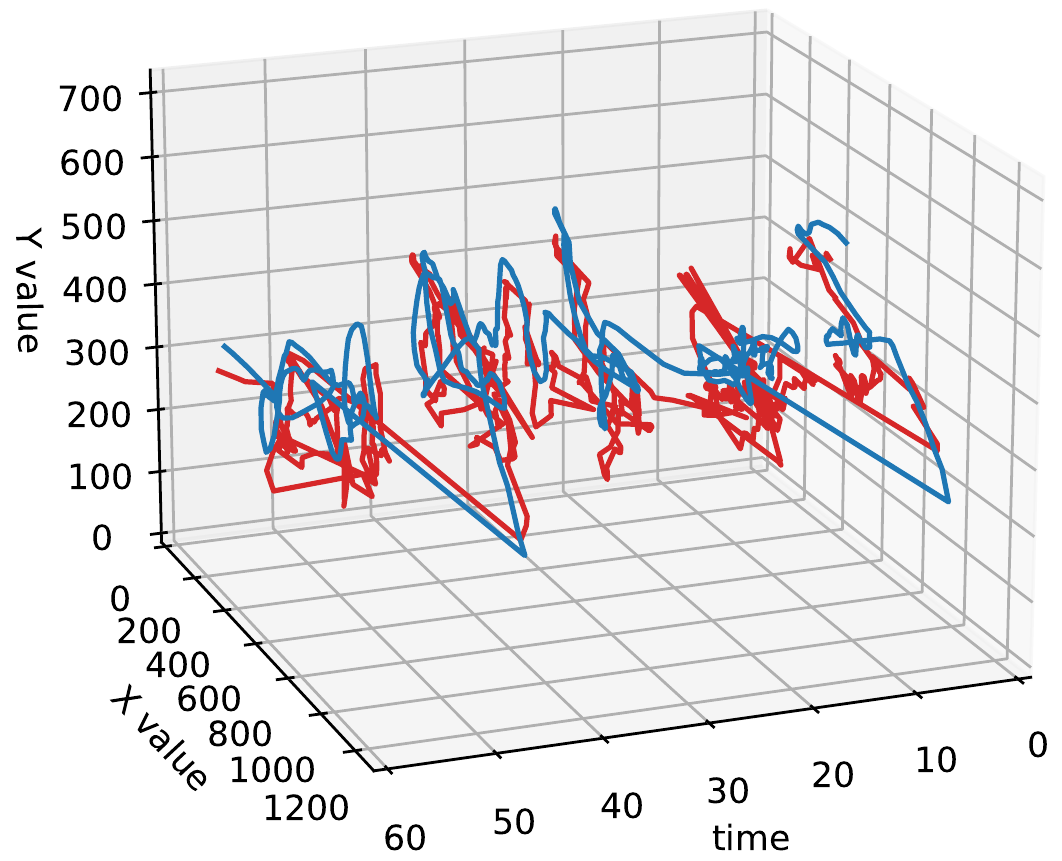}}
    \quad
    \hspace{-0.3cm}
  \subfigure[Video 9]{
    \label{fig:visual12}
    \centering
    \includegraphics[width=0.32\linewidth, height=115pt]{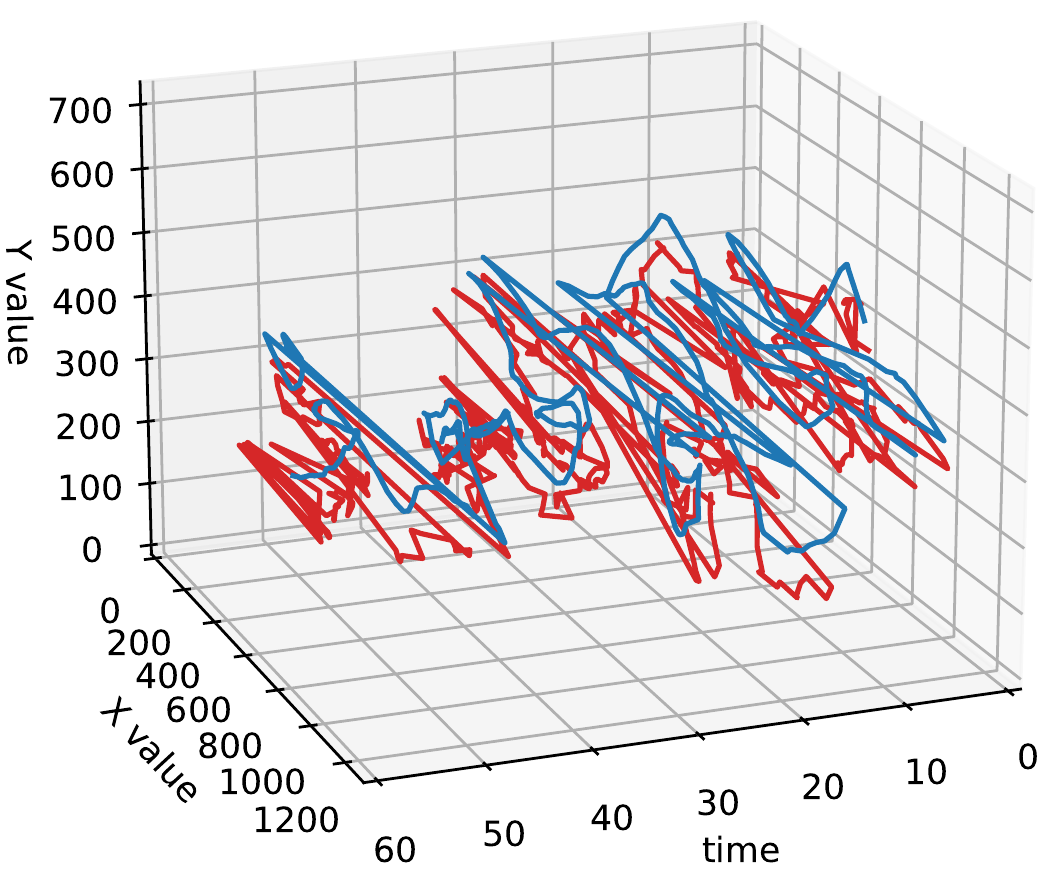}}
    \quad
    \hspace{-0.3cm}
\vspace{-0.5cm}
\caption{Gaze and head trajectories of one user with different types of videos}
\label{fig:visual_4}
\end{minipage}
\vspace{-0.5cm}
\end{figure*}

\begin{figure*}[!t]
\normalsize
\centering
\begin{minipage}[t]{\linewidth}
\vspace{0pt}
\centering
  \hspace{-0.5cm}
  \subfigure[Video 4]{
    \label{fig:visual13}
    \includegraphics[width=0.32\linewidth, height=115pt]{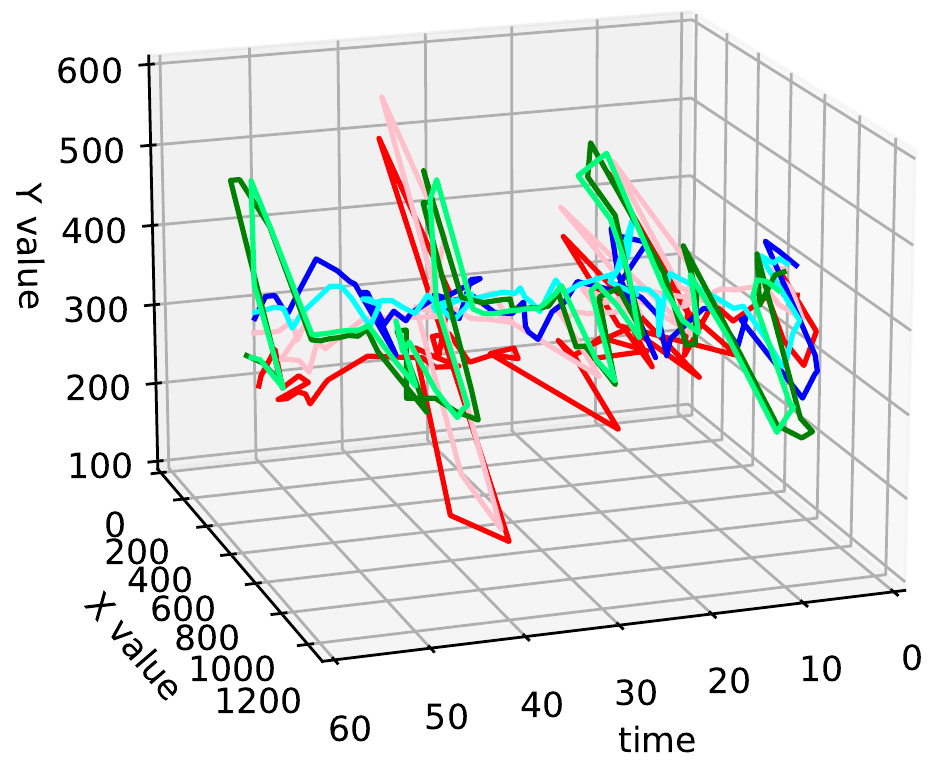}}
    \quad
    \hspace{-0.3cm}
  \subfigure[Video 5]{
    \label{fig:visual14}
    \includegraphics[width=0.32\linewidth, height=115pt]{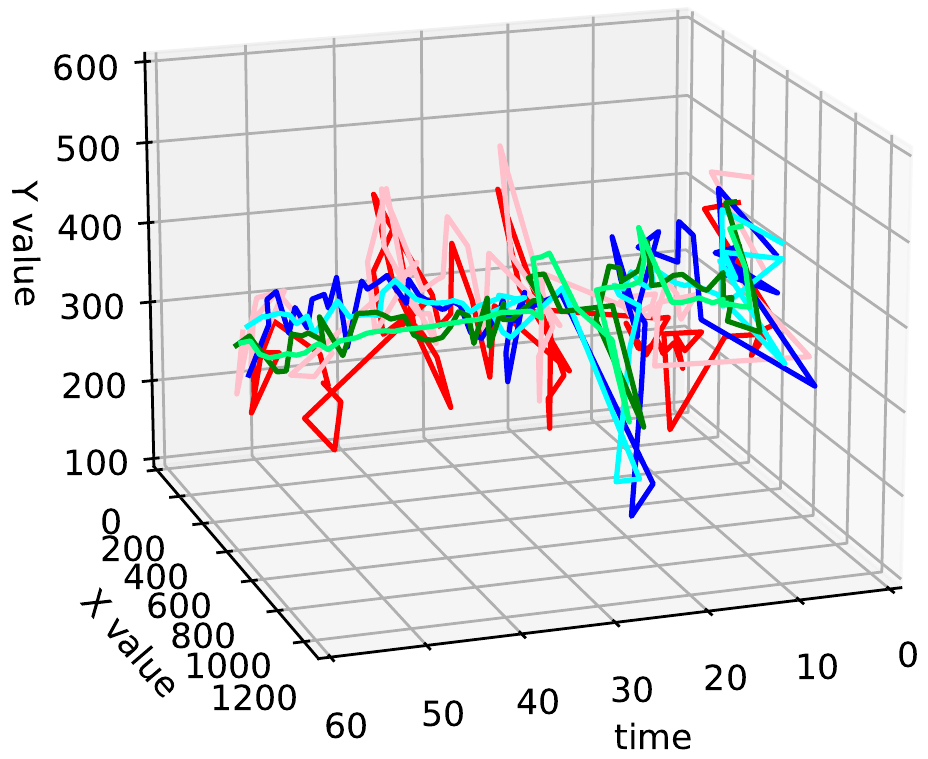}}
    \quad
    \hspace{-0.3cm}
  \subfigure[Video 6]{
    \label{fig:visual15}
    \includegraphics[width=0.32\linewidth, height=115pt]{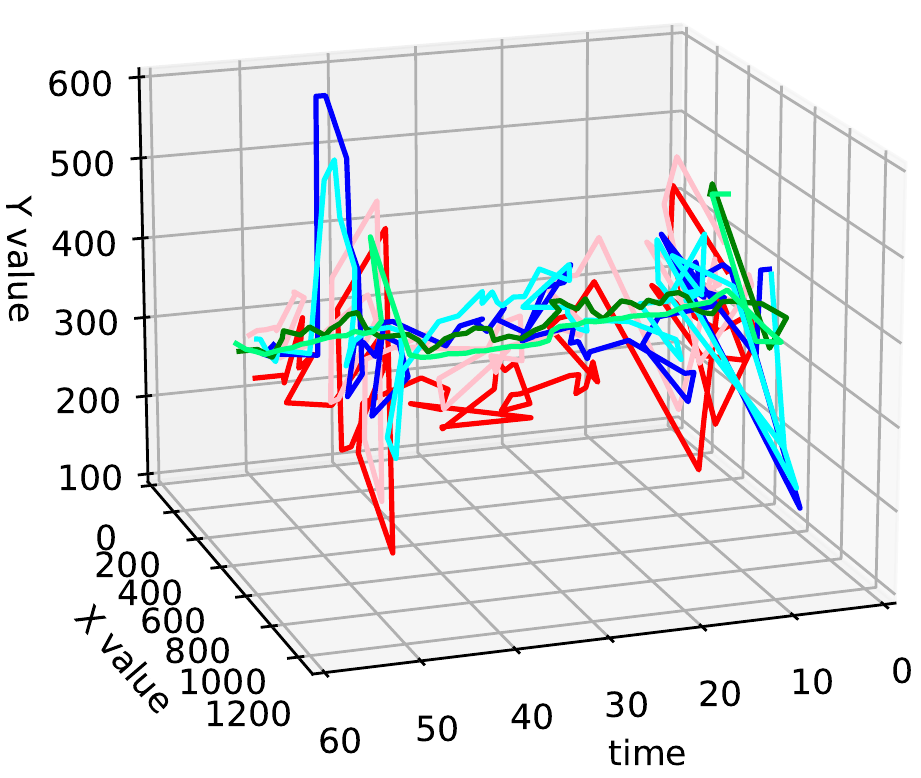}}
    \quad
    \hspace{-0.3cm}
\vspace{-0.5cm}
\caption{Gaze and head trajectories of three users with different types of watching motion}
\label{fig:figure5}
\end{minipage}
\vspace{-0.5cm}
\end{figure*}

\section{Dataset}
\label{sec:dataset}

We give details of our dataset in this section. Firstly, we show the procedure of data collection. Then the information from videos and users is presented. Finally, the structure of the data is listed.

\subsection{Collection Procedure}

The gaze and head tracking data are measured by VIVE Pro Eye\footnote{https://www.vive.com/us/product/vive-pro-eye/overview/}, a VR headset with precision eye tracking. VIVE provided SRanipal software development kit (SDK)\footnote{https://developer-express.vive.com/resources/vive-sense/} for developers to collect the gaze data. We develop a Unity\footnote{https://unity.com/} project based on OpenVR\footnote{https://github.com/ValveSoftware/openvr} and OpenXR\footnote{https://www.khronos.org/openxr/} to capture and save to a log file gaze data and head tracking data from the headset. Users are required to sit on a round stool. With the connecting line adaptive adjusted, the user can move the stool and rotate unconstrainedly, which enhances the sense of immersion when watching videos.

We clipped videos into three video collections with 9 videos. There are two seconds cooling frames between every two videos. Users can also pause the video to rest between videos.

Before the experiment, users are given enough time to adapt to the 3D environment. To get precise gaze data, users do gaze calibration, adjust the tightness of the headset and alter the position of the lens. Initially, they are oriented to the center of the video. After the program starts, they are free to look around and explore the environment to different degrees. When a video collection comes to an end, users are given time to rest. 

After watching, the API output the unit quaternion ($qx,qy,qz,qw$) to represent the rotation of the headset. For the convenience of usage, we transform data into two-dimensional points, where each point $(x, y)$ means the location of the user's head or gaze direction in the equirectangular projection of the video.

The unit vector $\mathbf{r}=(r_x,r_y,r_z)$ is calculated by unit quaternion from following equations: 
\vspace{-3pt}
\begin{equation}
\left[\begin{array}{c}
r_x \\
r_y \\
r_z
\end{array}\right]=\left[\begin{array}{c}
2 \times q x \times q z+2 \times q y \times q w \\
2 \times q y \times q z-2 \times q x \times q w \\
1-2 \times q x^{2}-2 \times q y^{2}
\end{array}\right]
\end{equation}
\vspace{-3pt}

By unit vector, we can calculate the angle between x-axis, y-axis shown in Figure \ref{fig:angle} by 
\vspace{-3pt}
\begin{equation}
    \varphi=\arctan \left(\frac{r_y}{r_x}\right), \theta=\arccos \left(\frac{r_y}{| r |}\right)
\end{equation}
\vspace{-3pt}

Then we can calculate $(x,y)$ by
\vspace{-3pt}
\begin{equation}
	y=\left\{
	\begin{array}{l}
		height - height \times \left(\cfrac{\theta} {2\pi}\right),\text{if }r_y \ge 0 \\
		height \times \left(\cfrac{\theta} {2\pi}\right),\text{if }r_y <  0
	\end{array}
	\right.
\end{equation}
\vspace{-3pt}
\begin{equation}
	x=\left\{
	\begin{array}{l}
		\cfrac{3width}{4} + sgn(r_x) \cfrac{\varphi} {2\pi} \times width, \text{ if }r_z>0  \\
		\cfrac{width}{2} - sgn(r_x) \cfrac{\varphi} {2\pi} \times width, \text{ if }r_z<0
	\end{array}
	\right.
\end{equation}\vspace{-3pt}
where $height$ and $width$ are the height and width of the video respectively, and $sgn()$ is the sign function. 

\subsection{Videos}

We collect videos from YouTube\footnote{https://www.youtube.com/}, and Table \ref{tbl:videos} shows the URL and content of the videos, where the full URL is the concatenation of "www.youtu.be/" and the string listed in the table.

While choosing videos, we consider the diversity of the videos based on the taxonomy we proposed. Table \ref{tbl:videos} roughly classifies videos into 9 categories, where each of them has 3 videos of different quality. Figure \ref{fig:videos} shows the quantitative metrics of each video and for convenience, we normalize each metric to $(0,1)$.  From the table and figure, we can find the videos we selected are diverse.

\subsection{Users}
After the users finish watching, he/she is asked to fill in a digital
questionnaire concerning the user’s gender, age, vision impairments level of familiarity with VR, and experience when watching 360° videos. Table \ref{tbl:users} presents the demographic profile of all users.

\section{Visualization}
\label{sec:vis}

This section shows some visualizations of our dataset, followed by some analyses.

\subsection{Direction on Moving Object}
The behavior of users can be analyzed by their head and gaze motion. The head and gaze direction is the point where the users are facing and looking when watching the 360° video. The point in spherical coordinate can be projected on the equirectangular image, as shown in Figure \ref{fig:visual_1} and Figure \ref{fig:visual_2}. To some extent, the moving of an object may drive the moving of the head and gaze. People are more likely to look at something dynamic. Figure \ref{fig:visual1} to Figure \ref{fig:visual3} are three moments from four seconds of Video 6. The ball-and-stick model represents the head and gaze direction. 

In the figures, red balls represent head direction; blue balls represent gaze direction. A light grey line links the head and gaze direction of each user. In Figure \ref{fig:visual1}, a couple of tangos are pivoting and gliding to the melodies. There is a tendency from left to right. Most of the users' heads and gaze direction are focused on the tangos. Their gaze direction is facing right compared to head direction. Figure \ref{fig:visual2} shows the users' head and gaze direction as the tangos move to the right. Many sticks move to the right. The length of the line is prolonged, showing the offset of gaze direction. Then, the tangos stayed on the right side. In Figure \ref{fig:visual3}, more sticks are on the right. Some are resting on the beginning of tangos. Based on these figures, it can be concluded that an object's moving direction drives the user's head or gaze to move in the same direction.

\subsection{Direction of Videos with Different ROI}
The content of the video determines the ROI of a user. In Figure \ref{fig:visual_1} to Figure \ref{fig:visual_3}, each row's content corresponds to videos with intensive, little dispersed, and very dispersed ROI. In Figure \ref{fig:visual4}, a bloodcurdling maumet is chasing the user. Most of the users are looking at the central picture, and the FoV is intensive. In Figure \ref{fig:visual5}, users ride a roller coaster with various cartoon scenes nearby. The ROI is a little dispersed, and some users look at the buildings on both sides. In Figure \ref{fig:visual6}, users fly across the sky and have a bird view of the sea. Users can enjoy the scene everywhere. Therefore, the ROI of video is very dispersed. The sticks are everywhere.

Figure \ref{fig:visual_3} and Figure \ref{fig:visual_4} present density maps and projection maps of Video 3, 6, and 9. Each density map contains the direction of three users in a video (red, blue, and green). The more time the user spends on a place, the deeper the color in this place. The projection map presents the x and y trajectories of head and gaze direction in 60 seconds (red represents the head, blue represents the gaze). Each second contains ten points between equal time intervals. Figure \ref{fig:visual7} to \ref{fig:visual9} indicate, videos with intensive ROI usually have only one central ROI. Videos with very dispersed ROI often have multiple ROI, for example, multiple objects or dynamic environments. users view content from different angles.  Figure \ref{fig:visual10} to Figure \ref{fig:visual12} show the gaze and head trajectories of different types of videos. The more dispersed the ROI, the faster and more dramatically the direction changes.

\subsection{Direction of Different Types of Users}
The moving of the head and gaze direction is also affected by the user's character. Trajectories from Figure \ref{fig:visual13} to Figure \ref{fig:visual15} are from videos 4,5 and 6 with similar dispersion of ROI. we plot the trajectories of head and gaze direction of three users with different watching motions:
Red for User 1, blue for User 2, and green for User 3. The deep color is head direction, the light color is gaze direction. By comparing the trajectories of different colors, the results show that the red one moves quickly in each video, the blue one moves averagely, and the green one moves slowly. Users can be classified into three groups.

\section{Pilot Study}
\label{sec:pilot}

In this section, we do a pilot study of head and gaze behavior for 360° videos. The study mainly focuses on content exploration, eye direction, and the relevance of head and gaze. We have some interesting findings.

\subsection{360° Content Exploration}

\begin{figure}[!t]
\normalsize
\centering
\begin{minipage}[t]{\linewidth}
\vspace{0pt}
\centering
  \hspace{-0.5cm}
  \subfigure[The whole videos]{
    \label{fig:explorationAll}
    \includegraphics[width=0.5\linewidth,trim = 30 20 30 30, clip]{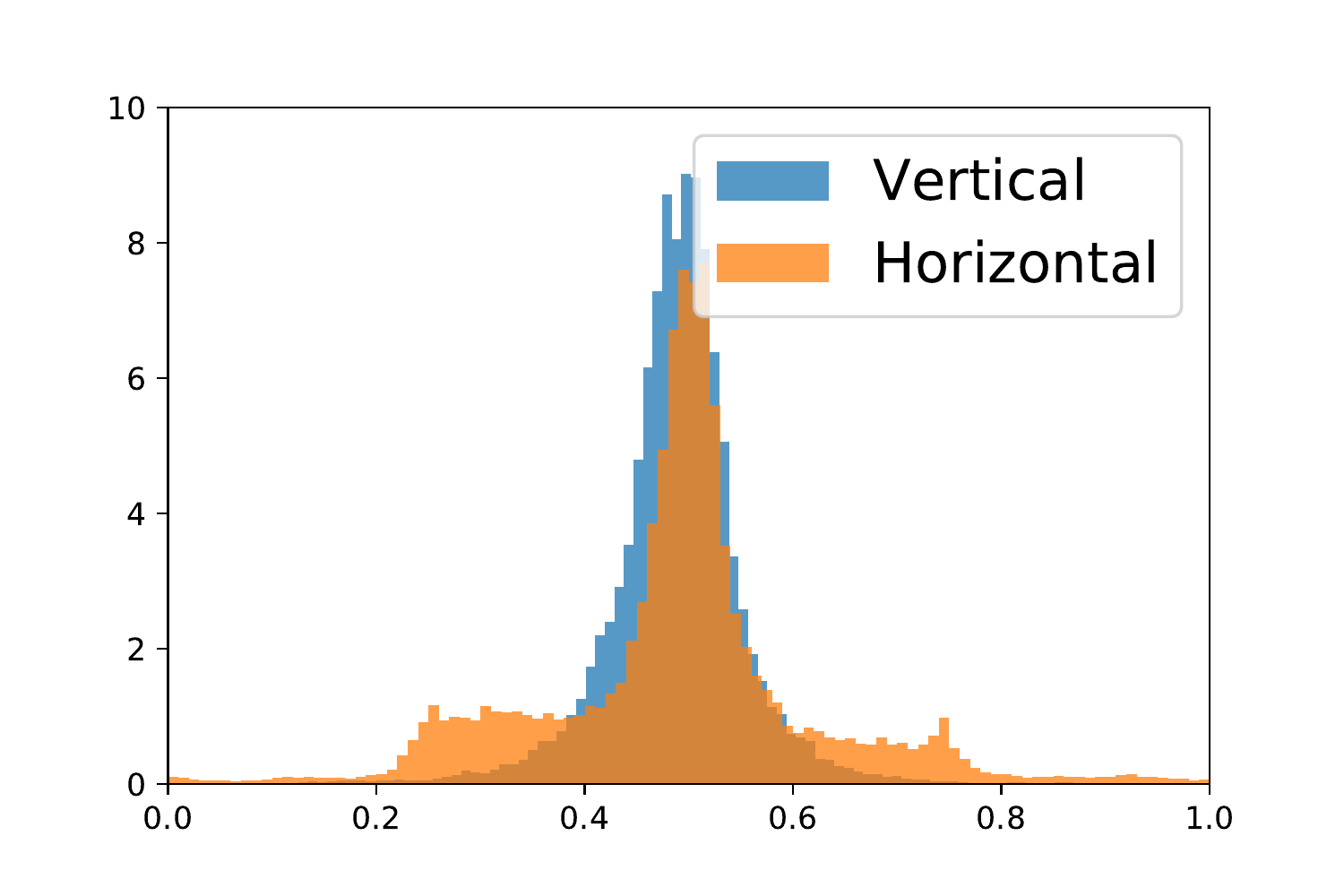}}
    \hspace{-0.3cm}
  \subfigure[The back of videos]{
    \label{fig:explorationPart}
    \includegraphics[width=0.5\linewidth,trim = 30 20 30 30, clip]{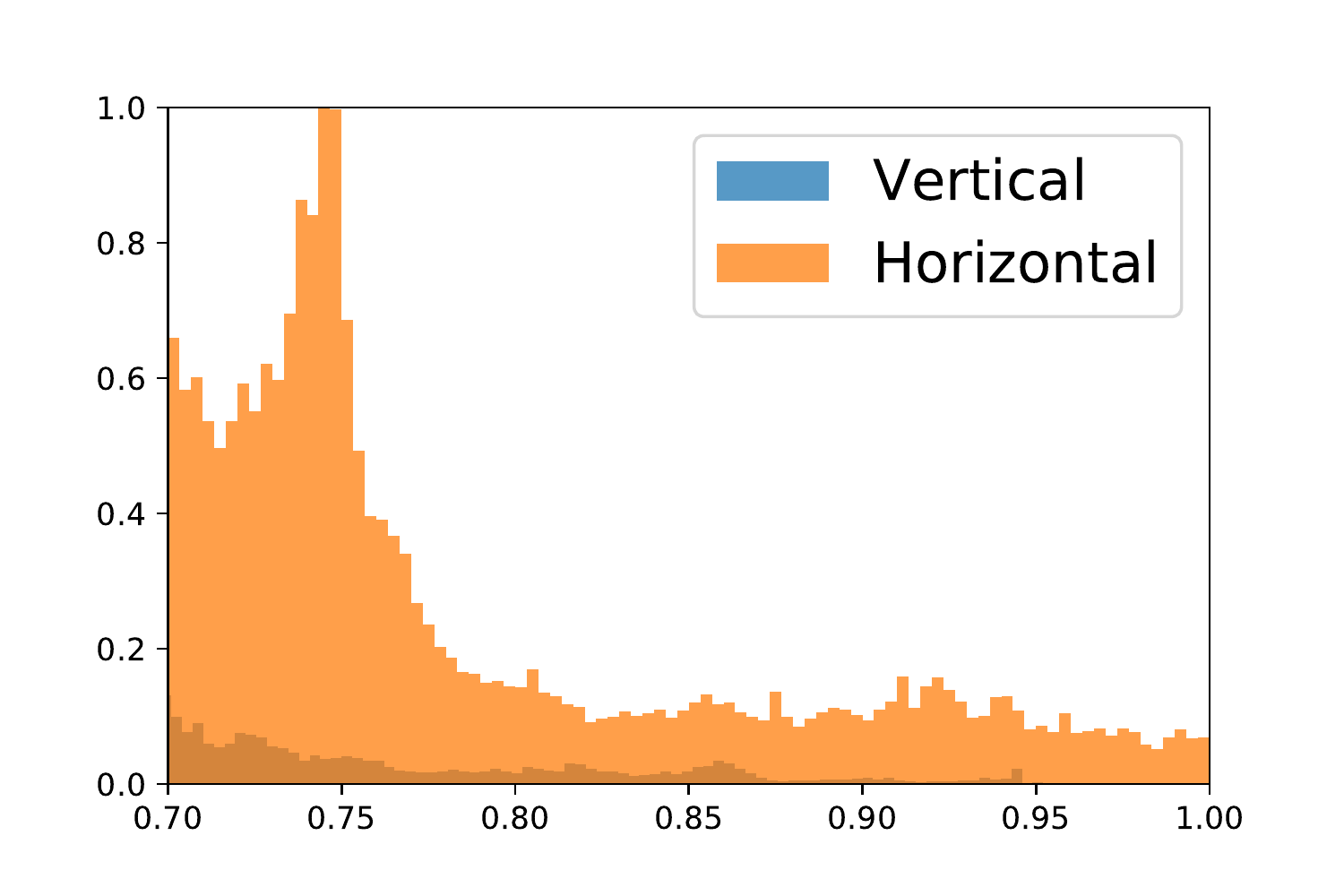}}
    \hspace{-0.3cm}
\vspace{-0.5cm}
\caption{Content exploration for all videos}
\label{fig:exploration}
\end{minipage}
\vspace{-15pt}
\end{figure}

The main difference between traditional videos and 360° videos is that users can freely explore the contents. It's interesting to study which part users watch more. So we count the number of data points from the horizontal axis and vertical axis.

Figure \ref{fig:explorationAll} shows the results for all videos and all users. Not surprisingly, \textbf{users mainly watch the center of the videos}.

But there are also differences between the horizontal axis and the vertical axis. Figure \ref{fig:explorationPart} shows the right 30\% of Figure \ref{fig:explorationAll} in detail. The right 30\% and the left 30\% compose the backside of the video. From the figure, we have that \textbf{users explore more horizontally than vertically}. And \textbf{the top and bottom of the videos are hardly ever been watched}.

\begin{figure}[!t]
\normalsize
\centering
\begin{minipage}[t]{\linewidth}
\vspace{0pt}
\centering
  \hspace{-0.5cm}
  \subfigure[Different Videos]{
    \label{fig:explorationdifferentVideos}
    \includegraphics[width=0.5\linewidth,trim = 30 20 30 30, clip]{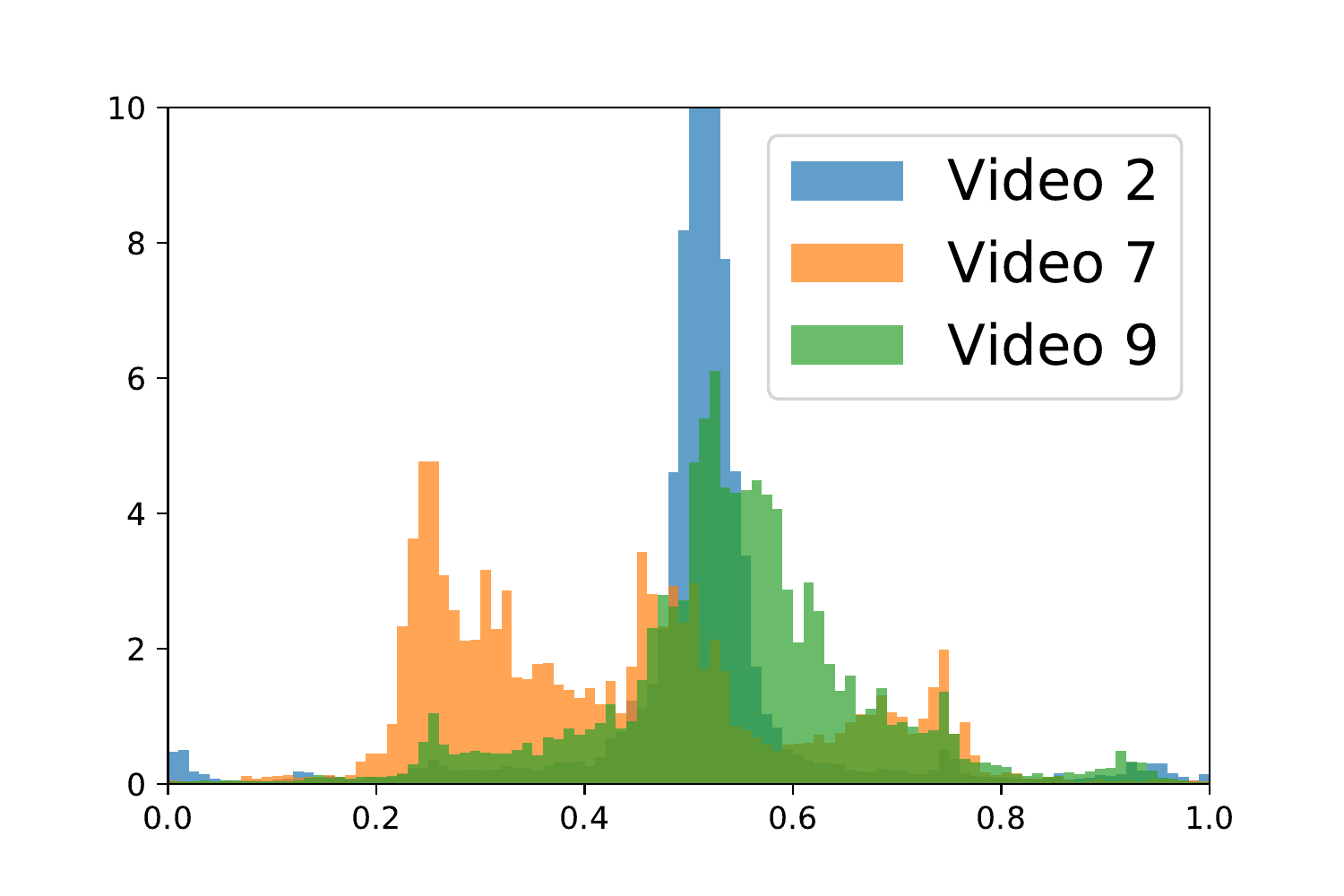}}
    \hspace{-0.3cm}
  \subfigure[Different Users]{
    \label{fig:explorationdifferentUsers}
    \includegraphics[width=0.5\linewidth,trim = 30 20 30 30, clip]{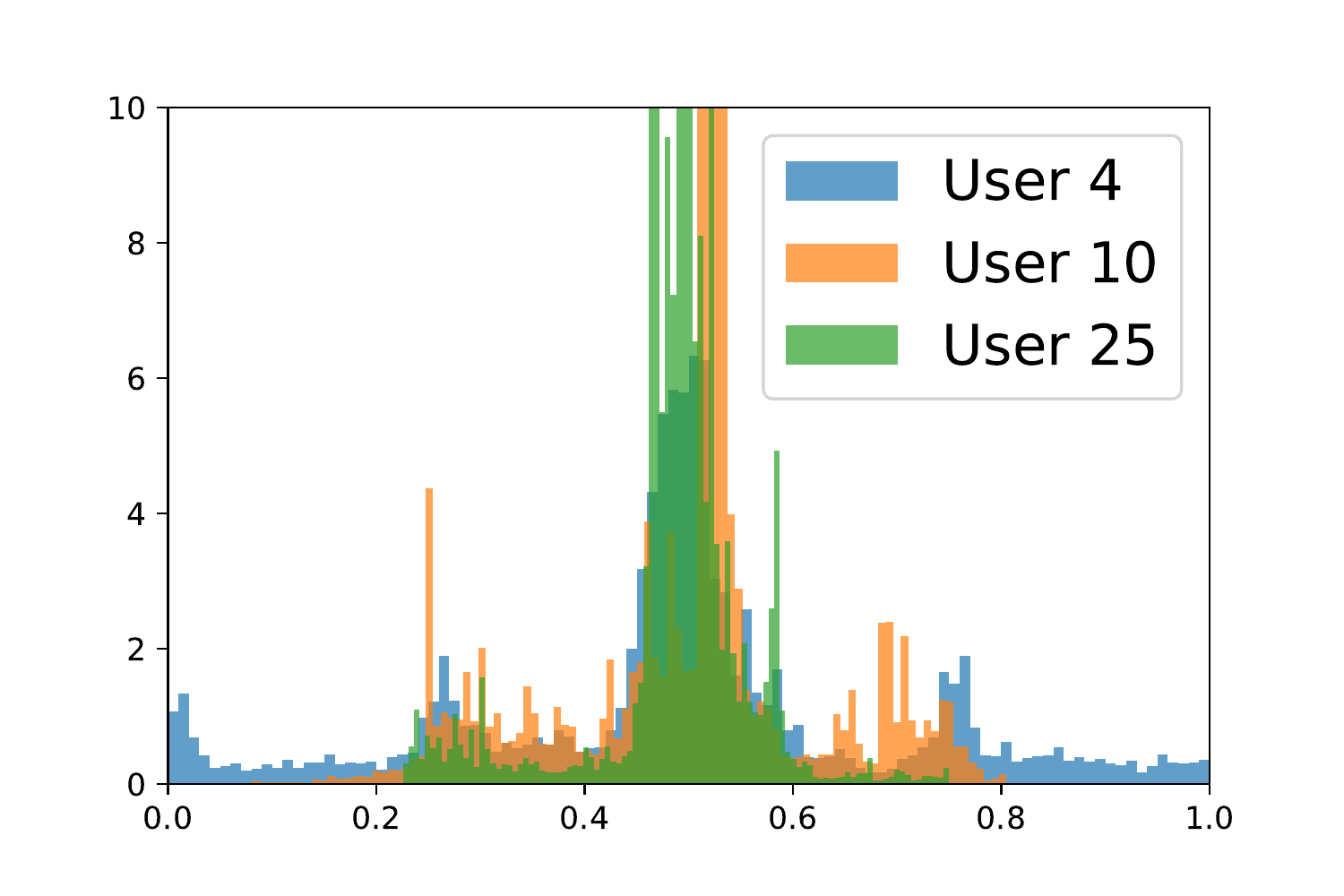}}
    \hspace{-0.3cm}
\vspace{-0.5cm}
\caption{Content exploration for different videos and users}
\label{fig:exploration2}
\end{minipage}
\vspace{-15pt}
\end{figure}

Figure \ref{fig:explorationdifferentVideos} shows the horizontal data of three different videos, with different dispersion of ROI. Figure \ref{fig:explorationdifferentUsers} shows the horizontal data of three different users. From these figures, we verify that \textbf{different videos let users focus on different parts} and \textbf{different users have different behavior patterns}.

\subsection{Eye Direction}

\begin{figure}[!t]
\centering
  \hspace{-0.2cm}
  \subfigure[All]{
    \label{fig:directionsAll}
    \includegraphics[width=0.32\hsize,scale=0.3, trim=90 40 90 40,clip]{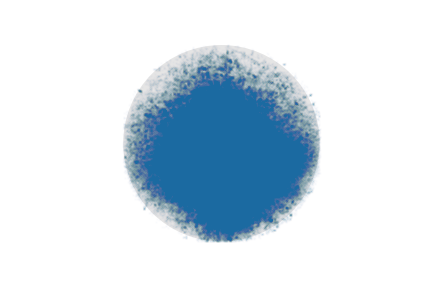}}
    \vspace{-6pt}
  \subfigure[User 1]{
    \label{fig:directionsUser1}
    \includegraphics[width=0.32\hsize,scale=0.3, trim=90 40 90 40,clip]{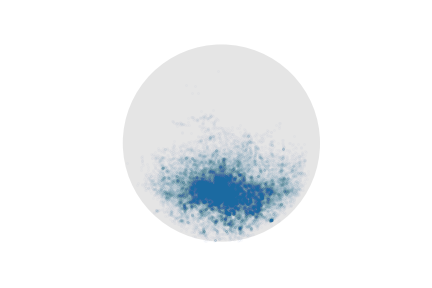}}
  \subfigure[User 24]{
    \label{fig:directionsUser24}
    \includegraphics[width=0.32\hsize,scale=0.3, trim=90 40 90 40,clip]{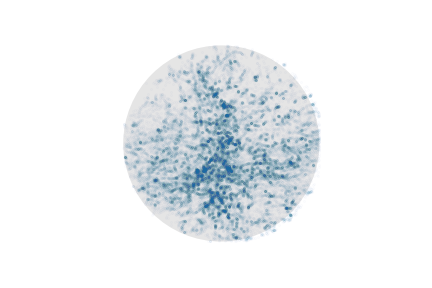}}
  \subfigure[Video 5]{
  \hspace{-0.2cm}
    \label{fig:directionsVideo5}
    \includegraphics[width=0.32\hsize,scale=0.3, trim=90 40 90 40,clip]{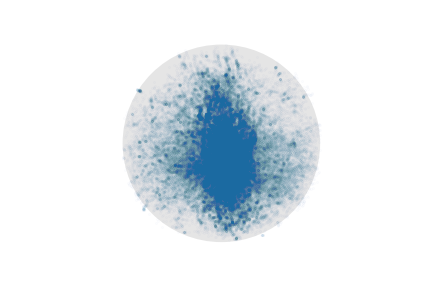}}
  \subfigure[Video 7]{
    \label{fig:directionsVideo7}
    \includegraphics[width=0.32\hsize,scale=0.3, trim=90 40 90 40,clip]{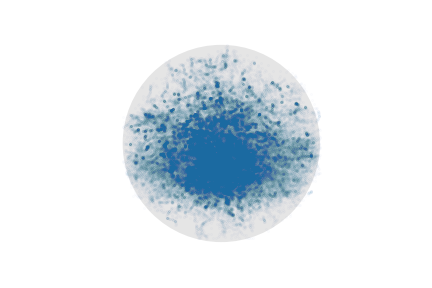}}
  \subfigure[Video 9]{
    \label{fig:directionsVideo9}
    \includegraphics[width=0.32\hsize,scale=0.3, trim=90 40 90 40,clip]{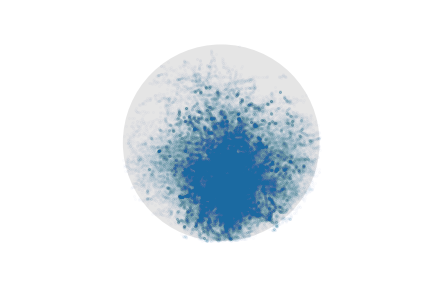}}
\vspace{-15pt}
\caption{The relative position of gaze and eye}
\label{fig:exploration2}
\vspace{-0.5cm}
\end{figure}

We are also interested in the relative position of head and gaze, which denotes your eye direction. Figure \ref{fig:directionsAll} to \ref{fig:directionsVideo9} present the heat-map of the positions of gaze, relative to the head center.

Figure \ref{fig:directionsAll} gathers points from all videos and all users. We find that \textbf{users hardly turn their eyes to four corners}.

Figure \ref{fig:directionsUser1} and \ref{fig:directionsUser24} show the heat-map of two different users and Figure \ref{fig:directionsVideo5} to \ref{fig:directionsVideo9} show the heat-map of three different videos. From these figures, we verify that \textbf{different videos leads different eye directions} and \textbf{different users have different behavior patterns}. User 1's eyes always focus on one point a little lower than the center but user 24 roll his/her eye frequently and irregular.

\subsection{Relevance of Head and Gaze}

\begin{figure}[!t]
\normalsize
\centering
\begin{minipage}[t]{\linewidth}
\vspace{0pt}
\centering
  \hspace{-0.5cm}
  \subfigure[Different Videos]{
    \label{fig:relationshipVideo}
    \includegraphics[width=0.5\linewidth,trim = 30 0 30 30, clip]{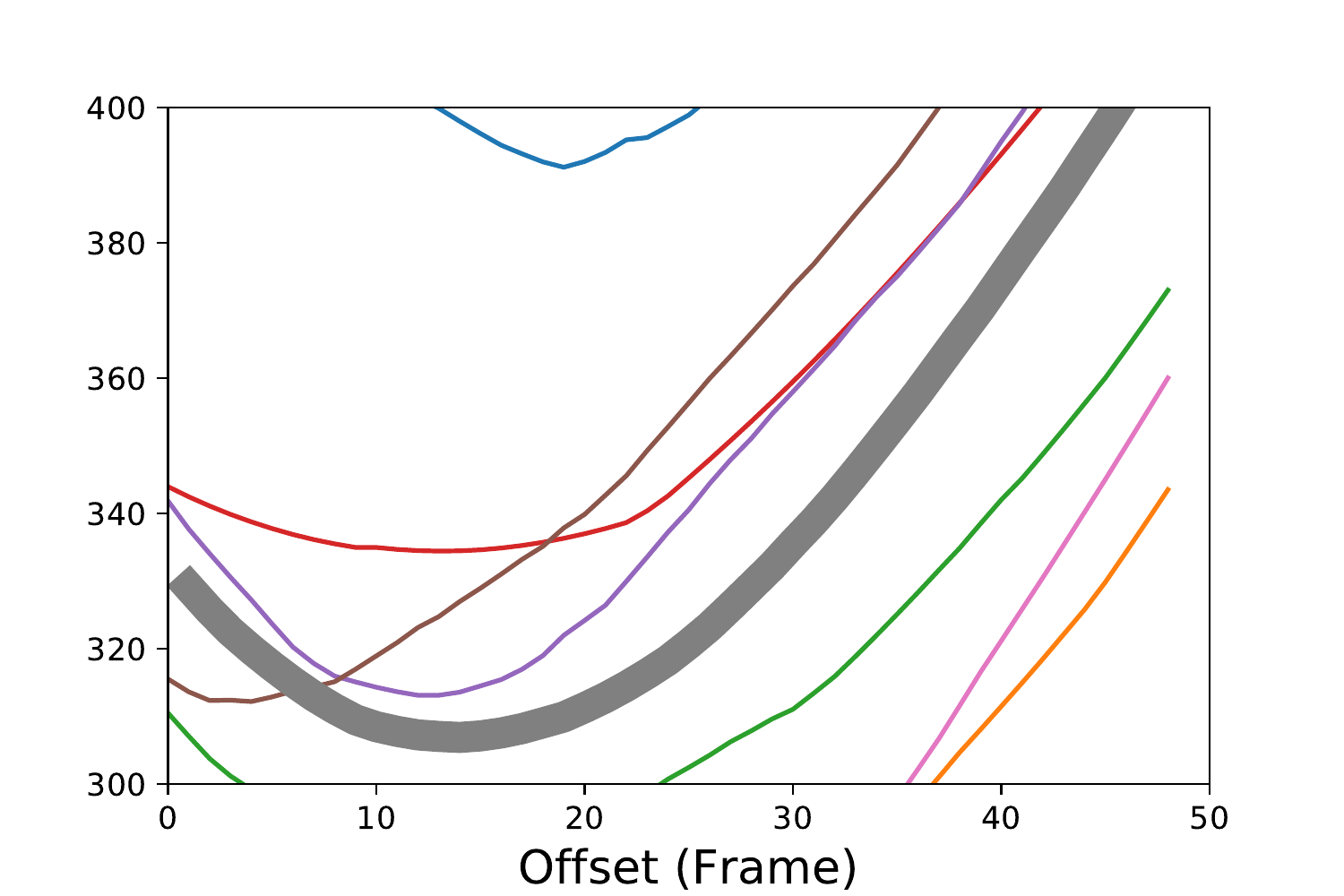}}
    \hspace{-0.3cm}
  \subfigure[Different Users]{
    \label{fig:relationshipUser}
    \includegraphics[width=0.5\linewidth,trim = 30 0 30 30, clip]{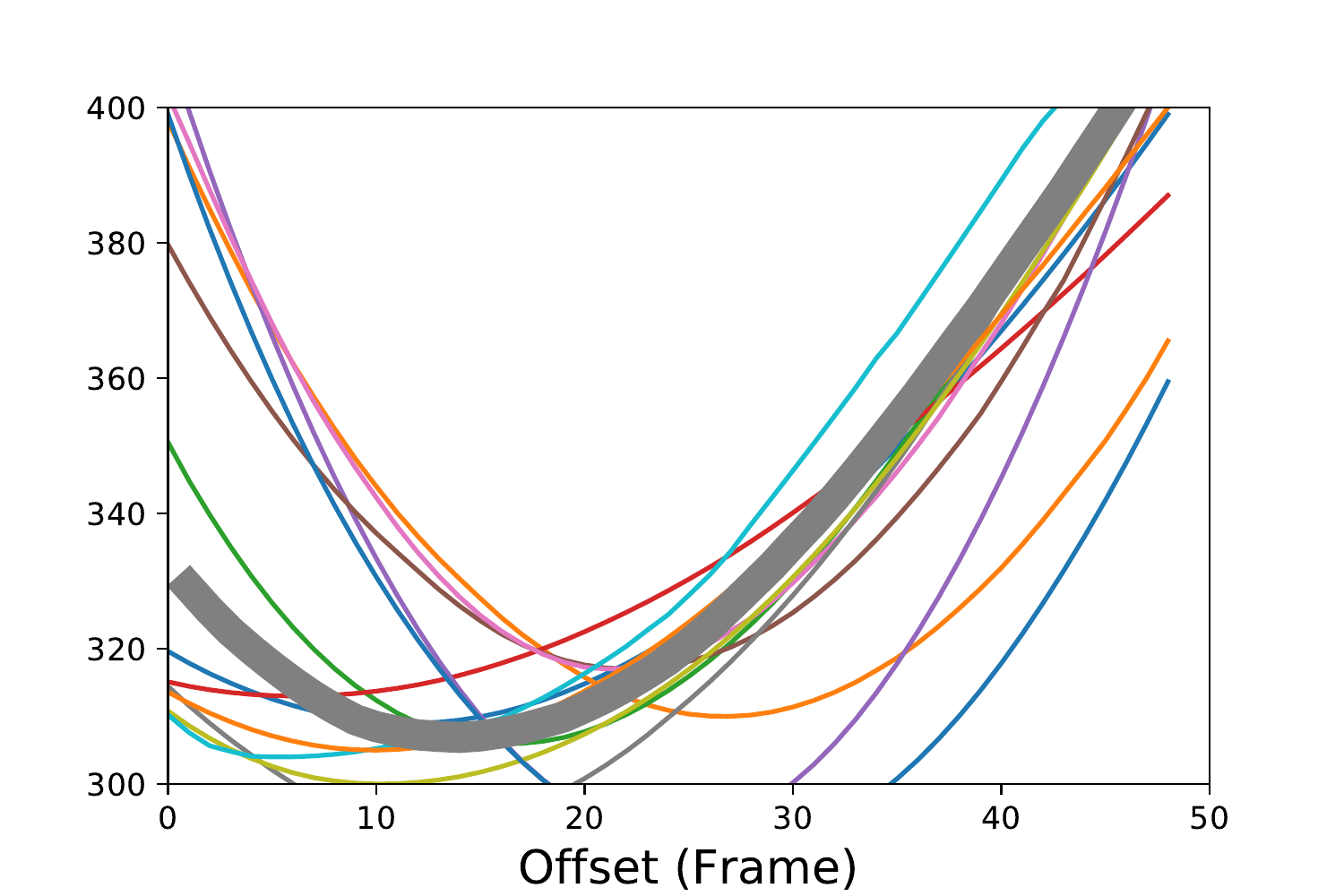}}
    \hspace{-0.3cm}
\vspace{-0.5cm}
\caption{MSE of head and gaze with different time offset}
\label{fig:relationship}
\end{minipage}
\end{figure}

\begin{figure}[tb]
	\centering
	\includegraphics[width=1\columnwidth, trim = 110 0 0 0,clip]{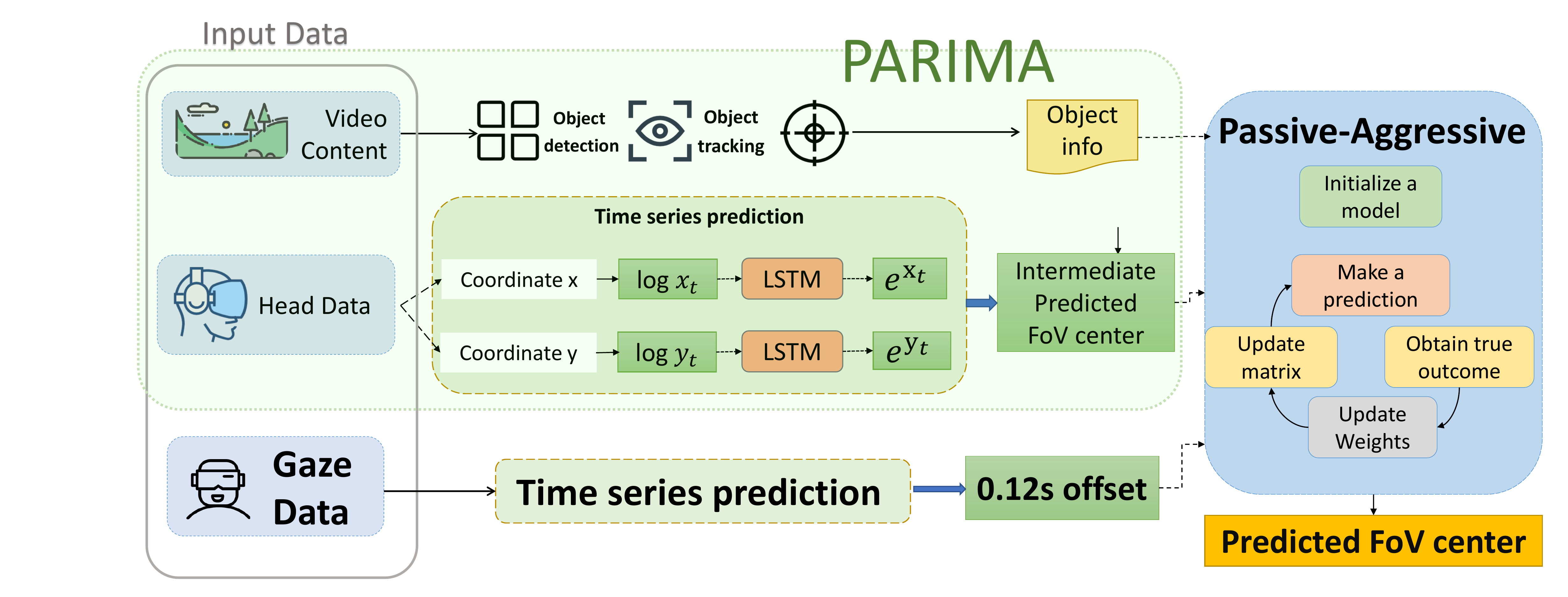}
	\vspace{-20pt}
	\caption{\label{fig:CFP}FoV prediction design}
	\vspace{-15pt}
\end{figure}

There is an interesting question: will users' heads movement follow gaze movement?

In this subsection, we calculate the deviation of head and gaze by adding an offset to verify this. We shift the gaze data forward from the timeline with different lengths, then calculate the Mean Square Error (MSE) between gaze and head. Figure \ref{fig:relationship} shows the results, where Figure \ref{fig:relationshipVideo} and Figure \ref{fig:relationshipUser} give the result of different videos and users respectively, and the thick grey line in the figures denotes the average result of all users and all videos.

The smaller MSE is, the similar the gaze and the head are. We surprisingly find that by adding a little time shift, the MSE declines. This means that the gaze data has more similarity with the head data in the future, but not now. The reason is probably that the user will turn his/her eye first, following with his head. 

From the experiments, we find that when the shift is 14 data points, i.e., about 0.12 seconds (the sampling frequency is 120Hz), the MSE is minimum. So we have that \textbf{user’s head direction will follow his/her gaze direction} and we can use current gaze data to assist to predict head position in about 0.12s future.

Due to the different behavior patterns of different users, from Figure \ref{fig:relationshipUser}, we can find that \textbf{for different users, the best prediction times are different, which vary from 0.08s to 0.2s}. In a real-world system, this property can also be used. If we have plenty of historical data for a user, we can take his/her own best time into consideration, and when a new user gets into the system, the average time of 0.12s can be applied.

\section{Gaze-Assisted 360° Video Streaming}
\label{sec:case}

In this section, we give a case of the application of our dataset in 360° video streaming. By leveraging gaze data, the efficiency of existing streaming systems can be highly improved.

\begin{table*}[t]
\begin{tabular}{|c|ccc|ccc|ccc|}
\hline
\multirow{2}{*}{\diagbox{Methods}{Metrics}}               & \multicolumn{3}{c|}{Unified Euclidean Distance}                                   & \multicolumn{3}{c|}{Manhattan Distance}                                           & \multicolumn{3}{c|}{Tile Accuracy}                                                \\ \cline{2-10} 
               & \multicolumn{1}{c|}{Only Head} & \multicolumn{1}{c|}{Gaze-Assisted} & Improvement & \multicolumn{1}{c|}{Only H.} & \multicolumn{1}{c|}{G.-A.} & Ipv. & \multicolumn{1}{c|}{Only H.} & \multicolumn{1}{c|}{G.-A.} & Ipv. \\ \hline
Polynomial Regression & \multicolumn{1}{c|}{0.45}      & \multicolumn{1}{c|}{0.32}       & 28.9\%            & \multicolumn{1}{c|}{3.42}          & \multicolumn{1}{c|}{2.54}              &  25.7\%         & \multicolumn{1}{c|}{58\%}          & \multicolumn{1}{c|}{69\%}              &  19.0\%     \\ \hline
LSTM                  & \multicolumn{1}{c|}{0.28}      & \multicolumn{1}{c|}{0.18}       &  35.7\%       & \multicolumn{1}{c|}{1.24}          & \multicolumn{1}{c|}{0.98}              &  21.0\%      & \multicolumn{1}{c|}{72\%}          & \multicolumn{1}{c|}{83\%}              &    15.2\%  \\ \hline
PARIMA                & \multicolumn{1}{c|}{0.13}      & \multicolumn{1}{c|}{0.09}       &  30.7\%           & \multicolumn{1}{c|}{0.63}          & \multicolumn{1}{c|}{0.48}              &  23.8\%      & \multicolumn{1}{c|}{92\%}          & \multicolumn{1}{c|}{96\%}              &     4.3\%   \\ \hline
\end{tabular}
\caption{FoV prediction performance}
\centering
\vspace{-20pt}
\label{tbl:fovpre}
\end{table*}

\subsection{Background and Related Works}
Streaming 360° videos requires much higher network bandwidth than traditional videos. To accommodate the high resource consumption, tile-based video encoding/streaming together with FoV-adaptive tile selection is proposed. Each frame of a video is split into different tiles and only tiles inside the user’s FoV are streamed at high quality. 

Accurate FoV prediction for 360° video is the key to tile-based adaptive streaming where many pioneer efforts have been made toward this goal. Many works\cite{DBLP:conf/bigdataconf/BaoWZRL16,DBLP:conf/mobicom/0001HXG18} use regression-based methodologies to predict the future FoV according to the historical trajectory, but they are not quite capable of capturing the inherent correlations. Further works\cite{DBLP:conf/mm/XieZG18,DBLP:conf/infocom/ZhangZBLSL19,DBLP:conf/nossdav/ChenHLWW20, DBLP:conf/vr/JinLW22} based on machine learning methods predict FoV more accurate. The above works are just based on historical trajectory and other works\cite{DBLP:journals/tmm/FanYHH20, DBLP:conf/mm/NguyenYN18} take video content into consideration, specifically, PARIMA\cite{DBLP:conf/www/ChopraCMC21} uses YOLOv3\cite{DBLP:journals/corr/abs-1804-02767} to detect the objects, then predict the FoV based on the track of objects.

\subsection{Design}
Here we show a potential application of our dataset to VR video streaming. Researchers can leverage gaze information to increase accuracy.

PARIMA\cite{DBLP:conf/www/ChopraCMC21} is the state-of-the-art method for FoV prediction. Based on PARIMA, we add gaze data to assist the FoV prediction.

Since we have already known that the best prediction time of gaze for the head is about 0.12s, the gaze information which is 0.12s forwards the predicted head data is used. To predict a further future, we first do a time series prediction of gaze, the specific method we leverage is the Long Short-Term Memory (LSTM) \cite{DBLP:journals/neco/HochreiterS97}.

We use the Passive-Aggressive Regression model\cite{DBLP:journals/jmlr/CrammerDKSS06}, which is an efficient online learning regression algorithm, to assign weights for raw PARIMA results and gaze information. The algorithm computes the mapping $\mathbb{R}^n \rightarrow \mathbb{R}$.

We run two Passive-Aggressive Regression models to predict $x$ and $y$ coordinates respectively. The equations for the prediction for next timeslot are given by:
\vspace{-5pt}
\begin{equation}
	\left\{
	\begin{array}{l}
		X_t = \theta_{0x} + \theta_{1x} X_t^{Gaze} + \theta_{2x} X_t^{Head} + \sum_{i=1}^{N_{obj}} \bm{\theta}_{3X} O_{Xit} \\
		Y_t = \theta_{0y} + \theta_{1y} Y_t^{Gaze} + \theta_{2y} Y_t^{Head} + \sum_{i=1}^{N_{obj}} \bm{\theta}_{3Y} O_{Yit}
	\end{array}
	\right.
\end{equation}
where $(X_t,Y_t)$ is the predicted centroid coordinate of FoV. $(X_t^{Gaze}$, $Y_t^{Gaze})$ is the corresponding gaze point, $(X_t^{Head},Y_t^{Head})$ is the intermediate result of head data, and $(O_{Xit}, O_{Yit})$ is the coordinates for the $i^{th}$ object detected by YOLO.

Besides PARIMA, we also use another two methods that are widely used, Polynomial Regression and LSTM. These methods are not as accurate as PARIMA, but their computation consumption is much lower than PARIMA and they do not use the information of video content. We simply input gaze information with a 0.12s offset as another variable to the model.

For each method, we use three metrics to evaluate it:
\begin{itemize}
    \item \textbf{Unified Euclidean Distance}: The Euclidean distance between the true FoV and the predicted one. We unify the width and height of the video to 1 for the convenience of comparison.
    \item \textbf{Manhattan Distance}: The Manhattan distance of the tile with the true FoV and the tile with the predicted one. Here we split the video into $8\times8$ tiles.
    \item \textbf{Tile Accuracy}: The accuracy of whether the method finds the correct tile.
\end{itemize}

Table \ref{tbl:fovpre} shows the results. From the table, we find that for all methods and all metrics, by adding gaze information, the prediction results are improved. For the state-of-the-art method, PARIMA, by integrating gaze information as an input of PA-regression, the Euclidean Distance between ground truth FoV and predicted FoV can be improved by about 31\%. In the situation of $8 \times 8$ tiles, the prediction accuracy can be improved from 92\% to 96\%.

\section{Other Applications}
\label{sec:app}

Our dataset has a variety of potential applications. This section provides some cases.

\subsection{User Identification}
Determine user's identification is an important task in 360° video, and the main usage is for authentication. Rogers et al.\cite{DBLP:conf/iswc/RogersWSV15} using classic statistical learning methods to identify users and Li et al.\cite{DBLP:conf/percom/LiAZXLG16} proved that simple head-movement patterns are robust against imitation attacks. Miller et al.\cite{miller2020personal} identify 95\% of users correctly when trained on less than 5 min of tracking data per person. Almost all works are based on head motion behaviors, and some of them\cite{DBLP:conf/icassp/MantyjarviLVMA05} leverage raw data collected from accelerometers in the device.

As our dataset contain the fine-grained and high sampling frequency of users’ head movements, it’s plausible to do user identification. Another potential point is that our dataset contains gaze information, which is more dynamic than head movements. It's reasonable that leveraging gaze information can establish a more accurate and robust system.

\subsection{Psycho Analysis}
Previous works proved that behaviors captured in tracking data can be associated with medical conditions such as ADHD\cite{DBLP:journals/computer/RizzoBSBKM04}, autism\cite{jarrold2013social}, and PTSD\cite{loucks2019you}. There is also growing literature on the use of tracking data to diagnose dementia\cite{cherniack2011not, werner2009use, tarnanas2013ecological}.

By adding another important dimension, gaze behavior, psychologists can conduct a more in-depth analysis. From the perspective of computer scientists, since the psychological conditions of users can be extracted from their behaviors, how to protect privacy in a VR system is an interesting application.

\subsection{Video Recommendation}
There are many complete works for video recommendation\cite{DBLP:conf/recsys/DavidsonLLNVGGHLLS10, DBLP:journals/jodsn/DeldjooECGPQ16}, while few works focus on 360° videos. Tsingalis et al.\cite{DBLP:conf/isccsp/TsingalisPP14} present a study on
YouTube recommendation graphs of 3D videos by using statistical relevance. Estrada and Simeone\cite{DBLP:conf/vr/EstradaS17} develop a recommendation system for physical object substitution in virtual reality.

Due to the difference between traditional videos and 360° videos, users' behaviors can be considered in the recommendation system. For example, based on the behaviors, we can better cluster users and excavate the relationship between users. Also, for a specific user, by analyzing his/her behavior, to find what kind of videos he/she would like is an interesting application. 

\section{Conclusions}
\label{sec:conclusion}
We mainly propose a large-scale dataset containing both head and gaze information, and further conduct comprehensive data analytics as well as a case application on gaze-assisted 360° video streaming. 

We first develop a quantitative taxonomy for 360° videos, which contains three objective technical metrics, i.e., camera motion, video quality, and the dispersion of region of interest (ROI). We then collect a 360° video dataset, which outperforms existing related datasets with rich dimensions (including both head and gaze), large scale (including 100 users and 27 videos), strong diversity (wide range across the taxonomy), and high frequency (120 Hz). A pilot study on users' behaviors is further conducted with some interesting findings. We next implement a case of the application based on our dataset in 360° video streaming. By leveraging gaze data, the efficiency of existing streaming systems can be highly improved. We also propose some other potential applications.

\begin{acks}
The work was supported in part by the National Key R\&D Program of China with grant No. 2018YFB1800800, by the Basic Research Project No. HZQB-KCZYZ-2021067 of Hetao Shenzhen-HK S\&T Cooperation Zone, by National Natural Science Foundation of China with Grant No.62102342, by Shenzhen Outstanding Talents Training Fund 202002, by Guangdong Research Projects No. 2017ZT07X152 and No. 2019CX01X104, and by the Guangdong Provincial Key Laboratory of Future Networks of Intelligence (Grant No. 2022B1212010001).
\end{acks}


\bibliographystyle{ACM-Reference-Format}
\bibliography{sample-base}

\appendix

\end{document}